\documentclass[aa,twocolumn,numberedappendix,twocolappendix,appendixfloats,tighten]{openjournal}

\usepackage{amsmath}
\usepackage[usenames, dvipsnames]{xcolor}
\usepackage{graphicx}
\usepackage{comment}
\usepackage{txfonts}
\usepackage{textgreek}
\usepackage{natbib,twoopt}
\usepackage{xspace}
\usepackage[colorlinks=true,allcolors=blue]{hyperref} 
\usepackage{soul}
\bibpunct{(}{)}{;}{a}{}{,} 
\makeatletter
\newcommandtwoopt{\citeads}[3][][]{\href{http://adsabs.harvard.edu/abs/#3}%
{\def\hyper@linkstart##1##2{}%
\let\hyper@linkend\@empty\citealp[#1][#2]{#3}}}
\newcommandtwoopt{\citepads}[3][][]{\href{http://adsabs.harvard.edu/abs/#3}%
{\def\hyper@linkstart##1##2{}%
\let\hyper@linkend\@empty\citep[#1][#2]{#3}}}
\newcommandtwoopt{\citetads}[3][][]{\href{http://adsabs.harvard.edu/abs/#3}%
{\def\hyper@linkstart##1##2{}%
\let\hyper@linkend\@empty\citet[#1][#2]{#3}}}
\newcommandtwoopt{\citeyearads}[3][][]%
{\href{http://adsabs.harvard.edu/abs/#3}
{\def\hyper@linkstart##1##2{}%
\let\hyper@linkend\@empty\citeyear[#1][#2]{#3}}}
\makeatother

\def\nc {n_{{\rm c}}}
\def\ec {e_{{\rm c}}}

\def\nci {n_{{\rm c},i}}
\def\eci {e_{{\rm c},i}}
\def\pci {P_{{\rm c},i}}
\def\gci {\gamma_{{\rm c},i}}
\def\pim {p_{i-1/2}}
\def\pip {p_{i+1/2}}

\makeatletter
\newcommand{\Rosdahl}{%
  \@ifundefined{cited@Rosdahl25}{%
    \citet{Rosdahl25}%
    \expandafter\gdef\csname cited@Rosdahl25\endcsname{}%
  }{%
    \hyperlink{cite.Rosdahl25}{R25}%
  }%
  \xspace
}
\makeatother

\usepackage[varvw]{newtxmath}
\renewcommand*\vec[1]{\ensuremath{\boldsymbol{#1}}}

\usepackage{orcidlink}

\usepackage{ulem}

\begin{document} 

\title{RAMSES-MCR: a consistent multi-group treatment of cosmic ray physics in momentum-space with the RAMSES code\vspace{-15mm}}

\author{Nimatou-Seydi Diallo$^{1*}$\orcidlink{0009-0007-5436-5450}
        Yohan Dubois$^{1\dag}$\orcidlink{0000-0003-0225-6387}
        Alexandre Marcowith$^2$\orcidlink{0000-0002-3971-0910}
        Joki Rosdahl$^3$\orcidlink{0000-0002-7534-8314}
        Beno\^it Commer\c{c}on$^3$\orcidlink{0000-0003-2407-1025}
}
\thanks{$^*$E-mail: \href{mailto:diallo@iap.fr}{diallo@iap.fr}}
\thanks{$^\dag$E-mail: \href{mailto:dubois@iap.fr}{dubois@iap.fr}}
\affiliation{$^1$ Institut d’astrophysique de Paris, Sorbonne Universit\'e, CNRS, UMR 7095, 98 bis boulevard Arago, 75014 Paris, France\\
$^2$ Laboratoire Univers et Particules de Montpellier, CNRS/Universit\'e de Montpellier, Montpellier, France\\
$^3$ Universit\'e de Lyon, Université Lyon1, ENS de Lyon, CNRS, Centre de Recherche Astrophysique de Lyon UMR5574, F-69230, Saint-Genis-Laval, France}
        

\begin{abstract}
    Cosmic rays (CRs) are known to play a key role in many astrophysical environments: they can modify shock dynamics, influence the thermochemistry and the ionization of the interstellar medium, regulate galaxy mass content by driving galactic winds, and be released by jets from active galactic nuclei. They also serve as important observational tracers through $\gamma$-ray emission, radio synchrotron, and secondary particle production.  
    Since CR particles follow power-law distributions in momentum space spanning many decades in energy, and because diffusion and radiative losses further shape these spectra, it is crucial to model spectrally resolved CRs in numerical simulations and to assess the impact of this modeling on gas dynamics and observational signatures.  
    We present a consistent multi-group spectral method in momentum space for CR protons called {\sc ramses-mcr} in the adaptive mesh refinement code {\sc ramses}, based on the two-moment formalism that evolves both CR energy and number density in momentum space, together with their associated flux. The modeled CR processes include advection, anisotropic/isotropic diffusion, streaming instability, Coulomb and hadronic losses, adiabatic changes, and feedback onto the gas.  
    We also show that the method can be naturally extended to CR electrons (e.g.~including synchrotron losses) and generalized to multiple CR species. The implementation is validated against a suite of standard multi-dimensional tests.
    We finally apply {\sc ramses-mcr} to the three-dimensional expansion of a supernova remnant including CRs with anisotropic diffusion and energy losses, and demonstrate how CR energy redistributes in a momentum-dependent manner and modifies the gas momentum during the snowplough phase.
\end{abstract}

\section{Introduction}

Cosmic rays (CRs) are high-energy particles whose energy density in galaxies is observed to be in rough equipartition with the kinetic, thermal, and magnetic energy densities of the interstellar medium (ISM) on galactic scales~\citep[e.g.][]{Boulares90,Beck05}. 
They therefore constitute a dynamically important and ubiquitous component of the ISM. 
Their presence is constrained both by direct detections in the Solar neighbourhood~\citep[e.g.][]{Adriani11,Aguilar14,Aguilar15,Cummings16,Gabici19}, and by non-thermal emission, including synchrotron radiation~\citep[e.g.][]{Condon92,Bell03,Helder12} and $\gamma$-ray emission~\citep[e.g.][]{Acero16,Abdollahi22}.

The bulk of Galactic CRs that significantly impact gas dynamics are protons. 
They are accelerated to non-thermal energies via the diffusive shock acceleration mechanism~\citep{Bell78,Blandford78,Caprioli15} in the shock wave of supernova remnants (SNR), as supported by their associated non-thermal emission, including $\gamma$-rays from pion decay~\citep{Helder09,Morlino12,Dermer13,Acero16}. 
Additional acceleration sites may contribute, such as active galactic nuclei, massive star winds, turbulent re-acceleration, or large-scale structure formation shocks, but SNRs are generally believed to dominate the CR budget in star-forming galaxies~\citep{Blasi13}.

CRs can significantly influence plasma dynamics due to their non-thermal nature and long lifetimes, and have therefore been proposed as an additional feedback agent in galaxies. 
Through their tight coupling to magnetic fields, CRs exert a pressure that contributes to the overall support of the gas~\citep{Hanasz04,Zweibel13}. 
In contrast to thermal energy, which can be rapidly radiated away (e.g.~$\sim 10$\,kyr for gas at a temperature $\sim10^6$\,K and a density $\sim10\,\mathrm{cm^{-3}}$), the bulk of CR proton energy is dissipated on much longer timescales (of order Myr under similar conditions), making CRs a comparatively long-lived energy reservoir.

CRs exchange energy and momentum with the thermal plasma through several processes. 
Coulomb and hadronic interactions transfer CR energy directly to the gas, while CR streaming along magnetic field lines excites Alfv\'en waves via the streaming instability~\citep{Kulsrud69}. 
The subsequent damping of these waves deposits heat into the thermal plasma. 
In addition, CR ionization regulates the thermo-chemistry of the ISM by setting the ionization rate in dense regions and driving ion–molecule reactions that control molecular formation~\citep{Grenier15,Padovani20}.
Because CRs have a softer effective equation of state (adiabatic index $4/3$) than the thermal gas ($5/3$), their pressure responds differently to compression and expansion, thereby modifying the thermodynamics of the gas.

A defining property of CR protons is their nearly collisionless nature. 
Because Coulomb interactions are weak at relativistic energies, their transport is governed not by ordinary collisional processes but by their interaction with magnetic fields. 
CRs gyrate around magnetic field lines and are scattered by magnetic fluctuations resonant with their Larmor radius, which isotropize their distribution and give rise, on macroscopic scales, to effective streaming and diffusion along the field lines. 
The origin of these scattering fluctuations remains debated: they may be self-generated by CRs through the streaming instability, or arise from a pre-existing MHD turbulent cascade~\citep[e.g.][]{Chandran00,Yan02,Jubelgas08,Zweibel17}, possibly exhibiting intermittent or patchy structures~\citep{Lazarian21,Lemoine23,Kempski23,Reichherzer25}. 
In either case, the resulting transport coefficients depend on particle energy, implying that CR propagation is intrinsically energy-dependent. 
On galactic scales, GeV CRs are typically characterized by diffusion coefficients of order $10^{28}\mbox{--}10^{29}\,\mathrm{cm^2\,s^{-1}}$, with an approximate power-law scaling with particle energy of index $\sim 0.3\mbox{--}0.6$~\citep{Evoli08,Evoli20,Trotta11,Bisschoff19}.
However, theoretical models and $\gamma$-ray observations of regions surrounding CR sources suggest that the diffusion coefficient can be significantly suppressed in their vicinity, down to values of order $10^{25}\mbox{--}10^{26}\,\mathrm{cm^2\,s^{-1}}$, due to enhanced self-generated turbulence~\citep[e.g.][]{Nava16,Nava19,Brahimi20,Marcowith25}.

Because CRs diffuse and stream efficiently while radiating energy slowly, they redistribute the energy injected by SNRs over kiloparsec scales. Although only a small fraction of the supernova (SN) explosion energy is transferred to CRs, their long lifetimes (to the point that CRs can significantly modify the SN injection of momentum~\citealp{Diesing18,Rodriguez22}) and efficient transport make them a relatively smooth, volume-filling feedback agent, in contrast to the more localized and rapidly cooling thermal energy deposited near SNRs. 
Indeed, hydrodynamical simulations that include CRs have demonstrated their important role in galaxy evolution. 
CRs can enhance mass outflow rates in galactic winds~\citep[e.g.][]{Uhlig12,Hanasz13,Booth13,Salem14,Girichidis16,Simpson16,Ruszkowski17,Armillotta24} in a non-trivial, mass-dependent manner~\citep{Jacob18,Dashyan20,Hopkins21}, modify wind thermodynamics~\citep{Girichidis18,Farcy22} as well as the properties of the circum-galactic medium~\citep{Buck20,Butsky20,Ji20,DeFelippis24,Farcy25}, regulate the gas reservoir available for star formation on small scales~\citep{Commercon19,Dubois19,Semenov21,Simpson23,Sampson25}, and ultimately alter global galaxy properties~\citep{Jubelgas08,Hopkins21CRgal,MartinAlvarez23,Rodriguez24,Bieri26}. 
CRs released by active galactic nuclei can likewise influence the dynamics of their inflated bubbles~\citep{Ruszkkowski17cluster,Ehlert18,Su26} and contribute to the heating and stability of the intra-cluster medium~\citep{Sijacki08,Guo08,Wang20,Su20,Su21,Beckmann22}. 
Taken together, these studies have established that CR feedback is fundamentally regulated by diffusive and streaming transport. 
Yet, most of these simulations rely on a grey approximation, treating CRs as a single-energy fluid with an effective transport coefficient. 
Given that both CR transport and losses depend intrinsically on particle energy, it remains unclear how spectral variations may modify the coupling between CRs and the gas. 

Indeed, not only CR transport but also CR energy losses are strongly energy dependent. 
Low-energy CR protons quickly lose energy through Coulomb and ionization interactions with the background plasma, while at higher energies (above $\sim1\,\rm GeV$) slow hadronic interactions dominate, producing pions that decay into $\gamma$-rays. 
As CRs propagate away from their sources, their spectrum evolves under the combined effects of energy-dependent transport and losses, leading to spatially varying spectral distributions.

This intrinsic spectral evolution calls for models that resolve the CR energy distribution rather than approximating it as a single fluid. 
Spectral CR transport solvers have long been developed in idealized frameworks without dynamical feedback, such as {\sc galprop}~\citep{Strong98}, {\sc dragon}~\citep{Evoli08}, {\sc picard}~\citep{Kissman14}, or {\sc usine}~\citep{Maurin01}, enabling detailed comparisons with CR observations. 
Early efforts also coupled spectral CR evolution to gas dynamics~\citep{Kang91,Jun99,Jones99,Miniati01,Jones05}. 
More recently, multi-group CR methods have been re-implemented in modern magneto-hydrodynamical (MHD) codes, allowing for dynamical feedback of spectrally resolved CRs onto the gas~\citep{Girichidis20}. 
These approaches have been used to study the impact of spectral evolution on galactic outflows (\citealp{Girichidis22} or~\citealp{Armillotta25}\footnote{In this work the feedback loop between CRs and MHD is not fully self-consistent as they use a semi-frozen approach where the transport of CRs is regularly post-processed and their new distribution are re-accounted for and injected in the MHD solver.}), to derive effective global diffusion coefficients~\citep{Girichidis24}, to investigate shock dynamics in galaxy clusters~\citep{Boss23}, and to confront transport scenarios with observations on galactic scales~\citep{Hopkins22,Hopkins22scenarios}. 
Similar methods have been used to study the evolution of CR electrons and their observable signatures~\citep[e.g.][]{Miniati01CRelectrons,Yang17,Vaidya18,Winner19,Vazza21,Ogrodnik21,Mukherjee21,Werhahn21,Linzer25,Ponnada25}.
However, only a limited number of studies have investigated the dynamical impact of fully coupled multi-group CR models in self-consistent MHD simulations, leaving open the broader question of how spectral effects modify CR transport, energy losses, and their feedback on gas dynamics across astrophysical environments.

From a numerical standpoint, CR diffusion poses a significant challenge. 
Its parabolic nature imposes a timestep constraint scaling quadratically with spatial resolution in explicit schemes, while streaming leads to even more restrictive constraints. 
Solutions include sub-cycling, implicit or semi-implicit methods~\citep{Sharma11,Dubois16,Dubois19,Pakmor16Implicit}, transverse flux limiters~\citep{Sharma07}, or regularization strategies for streaming~\citep{Sharma10}. 
Alternatively, taking the first moments of the CR distribution function in momentum space~\citep[e.g.][]{Thomas19,Hopkins22derivation} yields a two-moment formalism analogous to radiative transfer, resulting in a hyperbolic system with source terms. 
This approach, implemented in several codes including {\sc athena}~\citep{Jiang18}, {\sc arepo}~\citep{Thomas21}, {\sc gizmo}~\citep{Chan19}, and recently in {\sc ramses}~in \Rosdahl (\Rosdahl hereafter), has the advantage that the timestep scales linearly with resolution, at the cost of introducing a characteristic propagation speed equal to the speed of light, which can be artificially reduced~\citep{Gnedin01,Snodin06}.

In this paper, we introduce {\sc ramses-mcr}, a multi-group CR solver implemented in the adaptive mesh refinement MHD code {\sc ramses}~\citep{Teyssier02}, based on the two-moment numerical implementation of~\Rosdahl.
{\sc ramses-mcr} evolves spectrally resolved CRs on the mesh, consistently coupled to gas dynamics, and includes advection, anisotropic diffusion, streaming, radiative losses, and injection. 
In Section~\ref{section:method} we describe the numerical method. 
In Section~\ref{section:results} we present a suite of idealized tests validating the solver. 
We then apply it to the three-dimensional evolution of a SNR, a physically motivated and sufficiently complex environment to demonstrate the astrophysical relevance of the multi-group approach.
We summarize and conclude in Section~\ref{section:conclusion}.

\section{Method}
\label{section:method}
\subsection{Cosmic-ray magneto-hydrodynamics with {\sc ramses-mcr}}

The equations of ideal MHD used in the adaptive mesh refinement code {\sc ramses}~\citep{Teyssier02} are:
\begin{align}
&    \label{eq:mass} \frac{\partial \rho}{\partial t} + \vec \nabla. (\rho \vec{u})= 0 \, ,\\
&    \label{eq:momentum}\frac{\partial \rho \vec{u}} {\partial t} + \vec \nabla. \left(\rho \vec{u}\otimes \vec{u}+P-\frac{\vec B\otimes \vec B}{4 \pi}\right)= -\vec \nabla P_{\rm c}\\
&    \label{eq:energy}\frac{\partial e} {\partial t} + \vec \nabla. \left((e+P)\vec{u}-\frac{ (\vec B.\vec u)\vec B}{4 \pi}\right) = -\vec u.\vec \nabla P_{\rm c} +\mathcal{H}_{\rm c} + \Lambda_{\rm r} \, ,\\
&    \label{eq:magnetic} \frac{\partial \vec{B}} {\partial t} - \vec \nabla \times (\vec{u} \times \vec{B} )=0\, ,
\end{align}
where $\rho$ is the gas mass density, $\vec u$ is the gas velocity, 
$e=\rho u^2/2+e_{\rm th}+B^2/(8\pi)$ is the plasma energy, $P=P_{\rm th}+B^2/(8\pi)$ the plasma pressure, $\vec B$ is the magnetic field, $e_{\rm th}$ the gas thermal energy with ideal equation of state $P_{\rm th}=(\gamma-1)e_{\rm th}$ with $\gamma=5/3$ for ideal non-relativistic mono-atomic gas, 
$e_{\rm c}=\sum_i \eci$ is the total CR energy which is the sum of the individual groups of CR energies, each with its own equation of state $\pci=(\gci-1)\eci$ (and total CR pressure $P_{\rm c}=\sum_i \pci$), where $\gamma_{\rm c,i}$ ranges from $5/3$ to $4/3$ for respectively CR particles in the non-relativistic or relativistic regime, $\mathcal{H}_{\rm c}=\mathcal{H}_{\rm s}+\mathcal{H}_{\rm r}$ captures the heating of the gas by the streaming instability $\mathcal{H}_{\rm s}$ and by Coulomb and hadronic losses on CRs $\mathcal{H}_{\rm r}$, and $\Lambda_{\rm r}$ encapsulates the direct radiative loss and heat sources of the thermal component.
While momentarily dropping the source terms of the plasma energy equation, the set of MHD equations are solved with an approximate (7-waves) Harten-Lax-van Lee-Discontinuous Riemann solver~\citep{Miyoshi05}, and using the constrained transport algorithm for the induction equation~\citep[see][and references therein]{Teyssier06,Fromang06}.
Finally, the CR fluid follows transport equations with advection with the gas, diffusion, and streaming that we now detail further. 

Although CR electrons produce strong radiative signatures, they carry only a minor fraction of the total CR energy owing to their lower acceleration efficiency and rapid cooling. 
The CR pressure relevant for gas dynamics is thus dominated by ions, which we restrict to protons. 
We therefore focus on CR protons, noting that the formalism can be readily extended to other CR species.
We split the CRs into a multi-group decomposition in bins of their individual particle momentum $p$.
To update the distribution function $f(p)$ split into multiple groups of central momentum $p_i$ made of broken power-laws, one has to track both the CR energy density $\eci$ and the number density $\nci$~\citep{Miniati01}.
We follow the 2-moment equations of CR energy transport implemented in {\sc ramses}, as described in \Rosdahl, and based on the implementation by~\citealp{Jiang18}, which we extend to the transport of both the CR energy and number densities, for each CR bin of momentum $p_i$.
In fluid frame, those equations are:
\begin{align}
&    \label{eq:ncr} \frac{\partial \nci}{\partial t} + \vec \nabla . (\vec u \nci)+\vec \nabla.\vec F^\nci =  \mathcal{Q}^n_{i\pm}\, , \\
&    \label{eq:fncr} \frac{1}{\tilde c^2}\frac{\partial \vec F^\nci}{\partial t} + \vec \nabla \left(\frac{\nci}{3}\right) = -\sigma^n_{i}\vec F^\nci\, , \\
&    \label{eq:ecr} \frac{\partial \eci}{\partial t} + \vec \nabla . (\vec u \eci) + \vec \nabla.\vec F^\eci = \mathcal{Q}^e_{i\pm} \, ,\\
&    \label{eq:fecr} \frac{1}{\tilde c^2}\frac{\partial \vec F^\eci}{\partial t} + \vec \nabla \left(\frac{\eci}{3}\right) = -\sigma^e_{i}\vec F^\eci\, ,
\end{align}
where $\vec F^\nci$ and $\vec F^\eci$ for the CR number and energy density (fluid-frame) fluxes, respectively, $\tilde c$ is the reduced speed of light.
$\mathcal{Q}_{{\rm n},i\pm}$ and $\mathcal{Q}_{{\rm e},i\pm}$ capture the transfer of $\nci$ and $\eci$ (and associated losses and gains with bin $p_i$), respectively, between momentum bins $p_i$, and its two neighbors $p_{i-1}$ and $p_{i+1}$ due to radiative losses, adiabatic changes, streaming, second-order Fermi acceleration, and injection of fresh CRs.
This is the key aspect of the spectral method that we will detail in section~\ref{section:spectral}.
The interaction coefficient is split into parallel and perpendicular contributions with respect to the (unit vector) $\vec b=\vec B/\vert\vert\vec B\vert\vert$ field, i.e.: 
\begin{equation}
\label{eq:interaction_tensor}
\sigma^\phi_i = \sigma^\phi_{\parallel,i} \vec b\otimes\vec b +\sigma^\phi_{\perp,i} (\mathbb{I}-\vec b\otimes\vec b)\, ,
\end{equation}
where $\phi=\nc$ or $\ec$ (see~\Rosdahl for tests of the anisotropic and isotropic shapes of the tensor).
Each parallel and perpendicular terms are split into a diffusion and a pseudo-diffusive streaming contribution (only showing here the parallel terms):
\begin{align}
&\left(\sigma^\phi_{\parallel,i}\right)^{-1}=\left(\sigma^\phi_{{\rm d}\parallel,i}\right)^{-1}+\left(\sigma^\phi_{{\rm s}\parallel,i}\right)^{-1}\, ,\\
&  \left(3\sigma^\phi_{{\rm d}\parallel,i}\right)^{-1}=f_{{\rm d}\parallel,i}\kappa^\phi_i\, ,\\
&  \left(3\sigma^\phi_{s\parallel,i}\right)^{-1}=f_{{\rm s}\parallel,i}\frac{q_i}{3}\vert\vec u_{\rm A}\vert \frac{\phi_i}{\vert \vec \nabla_\parallel \phi_i\vert}\, ,
\end{align}
where $\vec u_{\rm A}$ is the Alfv\'en velocity, $\kappa_i^\phi$ is the diffusion coefficient, and $f_{{\rm d}\parallel,i}$ and $f_{{\rm s}\parallel,i}$ ($f_{{\rm d}\perp,i}$ and $f_{{\rm s}\perp,i}$) are the parallel (perpendicular) fractions for diffusion and streaming, respectively.
By default, our adopted values are $f_{{\rm d}\parallel,i}=f_{{\rm s}\parallel,i}=1$ and $f_{{\rm d}\perp,i}=f_{{\rm s}\perp,i}=10^{-6}$, unless otherwise specified.
We note that for the streaming instability, it is common to adopt $f_{{\rm s}\parallel,i}>1$ to capture the micro-physics of the damping of the self-excited Alfv\'en waves resulting in CR bulk propagation speeds larger than the Alfv\'en velocity (but not for the corresponding loss-gain term of the streaming stability, $\mathcal{H}_{\rm s}$, that still proceeds at the Alfv\'en speed). 

A notable difference with \Rosdahl is that the fluxes are now expressed in their fluid frame of reference\footnote{The comoving fluid frame fluxes $\vec F^\nci$ and $\vec F^\eci$ can be transformed into their lab frame expressions, respectively, $\vec F_{{\rm c,lab},i}^{n}$ and $\vec F_{{\rm c,lab},i}^{e}$, with $\vec F_{\rm c}^n=\vec F_{{\rm c,lab},i}^{n}-\vec u \nci$ and $\vec F_{\rm c}^e=\vec F_{{\rm c,lab},i}^{e}-\vec u(\eci+\pci)$.}
This formulation of the equations in fluid frame is i) necessary for the consistency of the spectral method under adiabatic transformations, ii) more rigorously derived from the Vlasov equations, and iii) is more accurate for small values of $\tilde c$ (see Appendix~\ref{appendix:frame_comp} for a comparison of the two formulations).

The CR equations \eqref{eq:ncr}–\eqref{eq:fecr} are solved using an operator-splitting approach.
The advective terms $\vec\nabla.(\vec u \phi_i)$ are treated together with the MHD solver.
Following \Rosdahl, the hyperbolic terms $\vec\nabla.\vec F^\phi_{{\rm c},i}$ and $\vec\nabla \phi_i/3$, as well as the right-hand-side source terms (including the CR pressure work in the gas momentum update), except for $\mathcal{Q}_{i\pm}^\phi$ (i.e.~the contribution from $-\sigma^\phi_i \vec F^\phi_{{\rm c},i}$) are solved using an explicit–implicit scheme (explicit for the hyperbolic terms and implicit for the source terms, see \Rosdahl for details). This step is advanced over a Courant-like timestep $\Delta t_{\rm cr,hyp}=\Delta x/\tilde c$, which is sub-cycled with respect to the MHD timestep $\Delta t_{\rm MHD}$.
Finally, the terms $\mathcal{Q}_{i\pm}^\phi$, which describe the transport of CR quantities in momentum space, are treated separately using a spectral timestep $\Delta t_{\rm cr,spe}$ and are sub-cycled with respect to $\Delta t_{\rm cr,hyp}$.

\subsection{The cosmic ray multi-group  spectral scheme}
\label{section:spectral}
\subsubsection{Discretization in momentum space and derivation of the four-moment cosmic ray fluid equations}

We derive the CR two-moment fluid equations (equations~\eqref{eq:ncr}-\eqref{eq:fecr}) in order to clearly identify the spectral steps of the method.

CRs can be represented by their gyrophase-averaged distribution function $f=f(x,p,\mu,t)$, where $x$ is the position, $p$ the particle momentum, and $\mu=\vec p.\vec b/p$ its pitch angle. 
Starting from the focused transport equations~\citep{Skilling71,Zank14}, one can describe the CR evolution with the following set of two equations taking the first two $\mu$-moments of the linear $\mu$-expansion of 
$f(x,p,\mu,t)=f_0(x,p,t)+3\mu f_1(x,p,t)$, where $f_0$ and $f_1=\langle\mu\rangle f_0$ are the isotropic and anisotropic parts of $f$ respectively~\citep{Thomas19,Hopkins22derivation}:
\begin{align}
\label{eq:f0} 
&\frac{\partial f_0}{\partial t} + \vec \nabla .(\vec uf_0)+ \vec \nabla. (v \vec b f_1) = \frac{1}{p^2} \frac{\partial}{\partial p} \left[p^2L(p)f_0\right] + j_0\, , \\
\label{eq:f1} 
&\frac{\partial f_1}{\partial t} + \vec \nabla. (\vec u f_1)+ v\tilde\nabla (f_0) = -\left[D_{\mu\mu}f_1+D_{\mu p}\frac{\partial f_0}{\partial p}\right] + j_1\, ,
\end{align}
with
\begin{align}
&\label{eq:bloss}L(p)=-\frac{{\rm d}p}{{\rm d}t}=L_{\rm r} + p\mathbb{D}:\nabla u + D_{p\mu}\frac{f_1}{f_0}+\frac{D_{pp}}{f_0}\frac{\partial f_0}{\partial p}
\end{align}
that encapsulates the gain/loss terms in momentum space and with scattering terms~\citep{Schlickeiser89}
\begin{align}
&D_{\mu\mu}=\bar \nu,\: 
D_{\mu p}=\chi\frac{p \bar u_{a}}{v} \bar \nu ,\: 
D_{p\mu}=\frac{p \bar u_{a}}{v} \bar \nu ,\: 
D_{pp}=\chi\frac{p^2 \bar u^2_{a}}{v^2} \bar \nu,
\end{align}
where $\tilde\nabla f_0=\vec b.\vec\nabla(\chi f_0)+\vec \nabla .((1-3\chi)f_0\vec b)$, $\mathbb{D}=\chi \mathbb{I}+(1-3\chi)\vec b\otimes\vec b$, and $\chi=(1-\langle\mu^2\rangle)/2$ correspond to anisotropic coefficients of the equations, and $v$ is the particle velocity. The scattering rate $\bar \nu=\nu_+ + \nu_-$ is the source of the diffusion in configuration space $x$. The signed Alfv\'en speed $\bar u_{\rm A}=u_{\rm A}(\nu_+-\nu_-)/(\nu_++\nu_-)$ corresponds to gyroresonant streaming instability terms, where $D_{\mu p}$ leads to a transport by streaming and $D_{p\mu}$ to a loss term.
The $L(p)$ term contains, in order of equation~\eqref{eq:bloss}, a term for: radiative losses ($L_{\rm r}>0$ for a loss term); adiabatic variations; streaming, and second-order Fermi acceleration.
The injection of CRs corresponds to the $j_0$ and $j_1$ terms.

\begin{figure}
    \centering
    \includegraphics[width=\columnwidth]{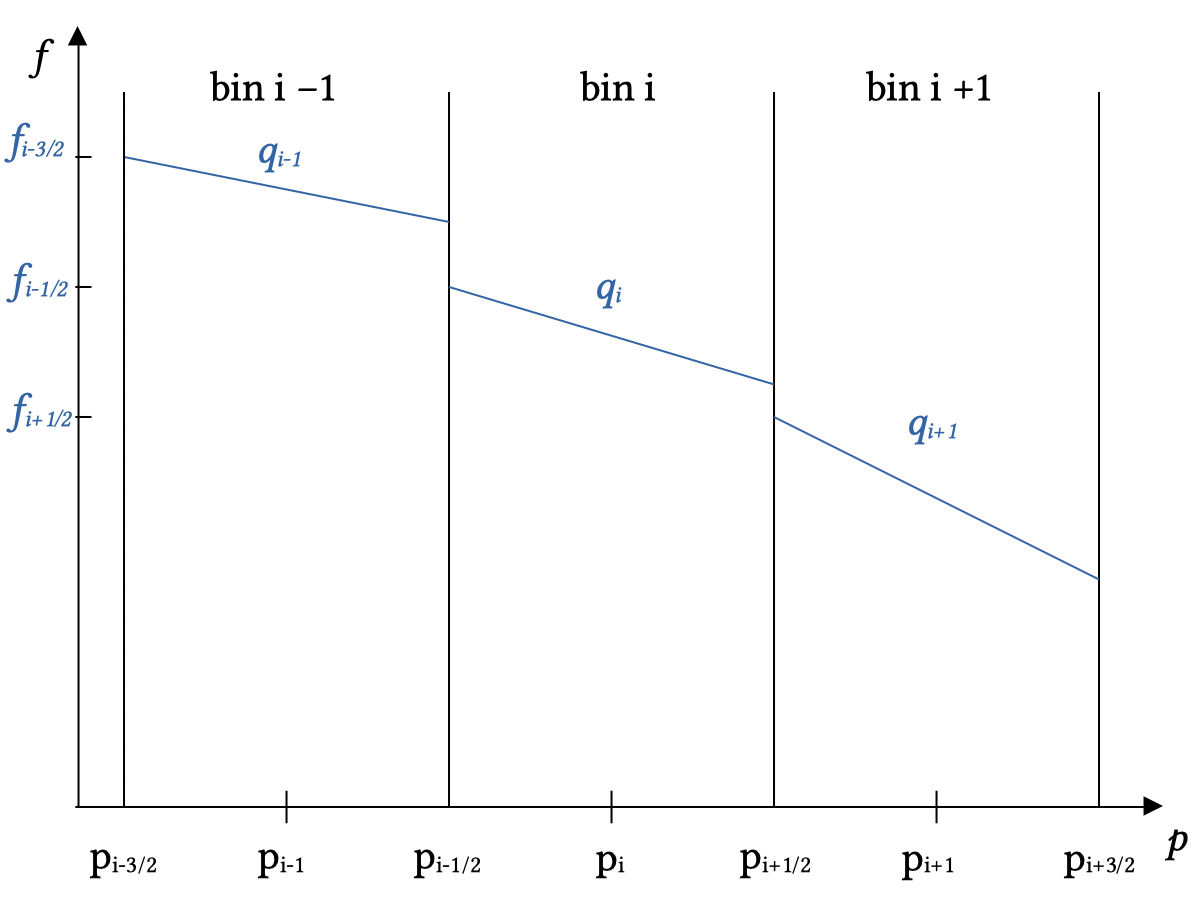}
    \caption{Illustration of the spectral discretization. Each momentum bin $i$ can be characterized by the parameters $(f_{i-1/2},q_i)$. The distribution function is not necessarily continuous at the interfaces.}
    \label{fig:discret_schema}
\end{figure}

The following lies on the foundations drawn by~\cite{Jones99} and~\cite{Miniati01} and later revisited by~\cite{Girichidis20}, and adapted to the two-moment method in~\citealp{Hopkins22}).
To represent and evolve the wide CR spectrum, we choose to discretize the spectrum into momentum bins centered around a momentum value $p_i$ that are equally spaced in logarithmic scale~\citep{Jones99}. 
With this discretization, illustrated in Fig.~\ref{fig:discret_schema}, we use a piecewise discontinuous distribution function.
Thus, for every momentum bin $i$, the distribution function goes as follows:
\begin{eqnarray}
    \label{eq:fp} 
    f_0(p)=f_{0,i-1/2}\,\left(\frac{p}{\pim}\right)^{-q_i}\,,
\end{eqnarray}
where $\pim\le p<\pip$ between the lower and upper bin limits, $f_{0,i-1/2}$ is the amplitude of $f_0$ at the right hand side of $\pim$, i.e.~for the given bin $i$ (the value of $f_0$ is discontinuous at the boundaries of the $p$-bins), and $q_i$ is the local inner-bin slope.

We can define the CR energy density $\eci$ and the number density $\nci$ for each bin:
\begin{align}
    \label{eq:ecr_i} 
    &\eci= 4\pi \int^{\pip}_{\pim}p^{2}T(p)f_0(p){\rm d}p\,,\\
    \label{eq:ncr_i} 
    &\nci= 4\pi \int^{\pip}_{\pim}p^{2}f_0(p){\rm d}p\, ,
\end{align}
where the kinetic energy of the CR particle is $T(p)=\sqrt{p^2c^2+m^2c^4}-mc^2$.
Therefore, it is possible to describe CR properties in a bin $p_i$ in an equivalent way by the pair of parameters $(\eci,\nci)$ or $(f_{0,i-1/2},q_i)$.
We can obtain the slope $q_i$ of the distribution function using the ratio
\begin{equation}
    \frac{\eci}{\nci}=\frac{\int^{\pip}_{\pim}p^{2-q_i}T(p){\rm d}p}{\int^{\pip}_{\pim}p^{2-q_i}{\rm d}p}
\end{equation}
and interpolating into precomputed tabulated values of $q_i$ versus $\eci/\nci$ for each bin $p_i$ of size $[\pim,\pip]$.
Once $q_i$ is known, the normalization $f_{i-1/2}$ is obtained analytically using equation~\eqref{eq:ncr_i}.

Although it is possible to follow the level of anisotropy of the distribution function, i.e.~$\langle\mu\rangle=f_1/f_0$ and obtain the anisotropic coefficients $\tilde\nabla f_0$ and $\mathbb{D}$ by using, for instance, an M1-closure (see e.g.~\citealp{Thomas22} for a comparison of different closures), in practice, we choose to simplify our equations by assuming a nearly 
isotropic\footnote{In most cases, this approximation ($\langle\mu\rangle=0$ and $\chi=1/3$) is sufficiently good as \begin{equation}
    \langle\mu\rangle=\frac{F^\phi}{\beta c\phi}\simeq \frac{\kappa} {cl}\simeq0.01 \frac{\kappa_{1\rm GeV/c}}{10^{28}{\rm cm^2 s^{-1}}} \left(\frac{p}{1{\rm GeV c^{-1}}}\right)^\delta \left(\frac{l}{10{\rm pc}}\right)^{-1}\, , \nonumber
\end{equation}
where $l=\phi/\vert\vec\nabla \phi\vert$ and $\delta\sim 0.5$. Hence, the approximation is going  to break down at about $1\,\rm TeV$ energies on scales of $l\sim 1\,\rm pc$.}
 distribution function ($\langle\mu \rangle\simeq0$, $\langle\mu^2 \rangle\simeq1/3$, $\chi\simeq1/3$, and $\mathbb{D}=\mathbb{I}/3$) but where we retain the equation on $f_1$ to still follow the flux functions for energy density $F^\eci$ and number density $F^\nci$:
\begin{align}
    \label{eq:fecr_i} 
&    F^\eci= 4\pi \int^{\pip}_{\pim}p^{2}T(p)vf_1(p){\rm d}p\,,\\
    \label{eq:fncr_i} 
&    F^\nci= 4\pi \int^{\pip}_{\pim}p^{2}vf_1(p){\rm d}p\, .
\end{align}

Additionally, the CR isotropic pressure is given by:
\begin{equation}
    \label{eq:pcr_i} 
    \pci= \frac{4\pi}{3} \int^{\pip}_{\pim}p^{3}c\beta(p)f_0(p){\rm d}p\,,
\end{equation}
with $\beta(p)=p/\sqrt{p^2+m^2c^2}=v/c$.
The corresponding $i$-th CR adiabatic index by $\gci=1+\pci/\eci$.
Again, we precompute and tabulate the values of $\gci$ as a function of $q_i$ to speed up the calculation of $\pci$ wherever required.

Equipped with these various quantities, we can integrate now equation~\eqref{eq:f0} over $4\pi p^2{\rm d}p$ and equation~\eqref{eq:f1} over $v4\pi p^2{\rm d}p$ to get the evolution of the CR number density and flux ($n_{\rm c}$,$F_{\rm c}^n$).
In the following, we neglect higher-order terms of order $v^{-2}$ arising from the moment expansion, except for those associated with the temporal evolution and relaxation of the flux.
Assuming bin-centred values of the integrands (i.e.~sufficiently narrow momentum bins), we obtain:
\begin{align}
&    \frac{\partial \nci}{\partial t} + \vec \nabla . (\vec u \nci) +\vec \nabla . (F^\nci\vec b)=\left[4\pi p^2 L(p) f_0\right]_{\pim}^{\pip} + j^n_{0,i}\, ,\\
\label{eq:truenumberflux}
&    \frac{1}{v^2}\frac{\partial F^\nci}{\partial t} +\vec b.\vec \nabla \left(\frac{\nci}{3}\right) = -\frac{\bar \nu_i}{v^2} \left[F^\nci-\frac{q_i}{3} \bar u_{\rm A} \nci \right]\, .
\end{align}
Similarly we can obtain the equations on energy and flux ($e_{\rm c}$, $F_{\rm c}^e$) by multiplying equations~\eqref{eq:f0} and~\eqref{eq:f1} by, respectively, $T(p)4\pi p^2{\rm d}p$ and $vT(p)4\pi p^2{\rm d}p$, and with the same set of approximations, after integration by parts:
\begin{align}
&    \frac{\partial \eci}{\partial t} + \vec \nabla . (\vec u \eci) +\vec \nabla.(F^\eci\vec b)= \left[4\pi p^2 L(p) T(p) f_0\right]_{\pim}^{\pip} \nonumber\\
&        - 4\pi \int^{\pip}_{\pim}p^2 v L(p) f_0 {\rm d}p +j^e_{0,i}\, ,\\
\label{eq:trueenergyflux}
&    \frac{1}{v^2}\frac{\partial F^\eci}{\partial t} +\vec b .\vec \nabla \left (\frac{\eci}{3} \right)= -\frac{\bar \nu_i}{v^2}\left[F^\eci-\frac{q_i}{3}\bar u_{\rm A} \eci\right]\, .
\end{align}

Therefore a spectral description of the CR system leads to the equations of evolution for respectively $\nci$ and $\eci$ (equations~\eqref{eq:ncr} and \eqref{eq:ecr}), where we use the reduced speed-of-light approximation $\tilde c$ ($<v$, which is a good approximation as long as $\tilde c$ and $v$ are sufficiently large compared to any of the transport velocities by advection, streaming or diffusion), and where we used 
$\kappa_\parallel=v^2/(3\bar \nu)=(3\sigma)^{-1}$.
We replaced the transport part of streaming affecting the flux equations ($-q_{i}\bar u_{\rm A}\phi_i/3$ in equations~\eqref{eq:truenumberflux} and~\eqref{eq:trueenergyflux}) by an additional factor in the interaction coefficient $\sigma$ that we finally write in the tensor form shown in equations~\eqref{eq:interaction_tensor} to allow for an arbitrary level of anisotropy for the diffusive and streaming  components of the flux.
Although this is not equivalent mathematically, this modification reduces to the same CR flux in the steady-state limit $\tilde c^{-2}\partial_tF_{\rm c}\rightarrow0$ and offers better numerical stability in various test cases as previously shown for the grey implementation of the method in \Rosdahl.

It becomes now apparent what the terms $\mathcal{Q}^e_{i\pm}$ and $\mathcal{Q}^n_{i\pm}$ in equations~\eqref{eq:ncr} and~\eqref{eq:ecr} are:
\begin{eqnarray}
    \mathcal{Q}^n_{i\pm}&=&\left[4\pi p^2 L(p) f_0(p)\right]_{\pim}^{\pip}+j^n_{0,i}\, , \\
    \mathcal{Q}^e_{i\pm}&=&\left[4\pi p^2 L(p) T(p) f_0(p)\right]_{\pim}^{\pip} \nonumber \\
\label{eq:qe}
    &-& 4\pi \int^{\pip}_{\pim}p^{2}\beta(p)c L(p)f_0(p){\rm d}p +j^e_{0,i}\,,
\end{eqnarray}
with $L(p)=-{\rm d}p/{\rm d}t$
corresponding to the total momentum variation over time by radiative losses, adiabatic changes, streaming and Fermi II losses.
The first term of the right hand side of these equations correspond to the flow of CR particles in $p$-space through the $p$-bin interfaces due to each of the $L(p)$ process.
The second term of the right hand side of equation~\eqref{eq:qe} corresponds to the loss/gain rate for the given $L(p)$ process.
The handling of these two contributions from $L(p)$ is detailed in section~\ref{section:spectral_update}.

Concerning the second term of the right hand side of equation~\eqref{eq:qe}, it is immediately recognizable that for $L(p)=p\vec \nabla.\vec u/3$, this term is simply $-\pci\vec \nabla.\vec u$.
In addition, this integral for $L(p)$ is: 
\begin{equation}
    \label{eq:stream_loss}
    L_{\rm s}=f_0^{-1}(D_{p\mu}f_1+D_{pp}\partial_p f_0)=p\frac{\bar\nu_i}{v^2}\bar u_{\rm A}\left(\frac{F^\eci}{\eci}-\frac{q_i}{3}\bar u_{\rm A}\right)\, ,
\end{equation}
which corresponds to the classical loss term (corresponding to the gain term  $\mathcal{H}_{\rm s}$ for the plasma) of the streaming instability $-(\bar\nu_i/v^2)3(\gci-1)\bar u_{\rm A}(F^\eci-q_i\bar u_{\rm A}/3)$ (using $p$-bin centric approximations), which further simplifies into $+\bar u_{\rm A}\vec b.\vec\nabla \pci$ in the steady-state limit of the flux (see Appendix~\ref{appendix:extra_equations} for the complete expressions of the CR quantities and their associated fluxes, and their reduced forms in appropriate limits).

For the streaming instability term, we reformulate $L_{\rm s}$ into $L_{\rm s}=p\sigma^\eci \vec u_{\rm s}.\vec  F^\eci/\eci$ for consistency of our reformulation of the streaming transport into the interaction coefficient $\sigma$  with $\vec u_{\rm s}=-\vec u_{\rm A}{\rm sign}(\vec b.\nabla \eci)$ and allowing for a non-purely anisotropic propagation of CRs along \vec b.

\subsubsection{Radiative losses}
Since we only consider a population of CR protons, their radiative loss term can be decomposed into Coulomb, ionization, and hadronic interactions $L_{\rm r}=L_{\rm r,C}+L_{\rm r,ion}+L_{\rm r,h}$.

For the loss rate resulting from Coulomb interactions between CR ions and electrons, we adopt the approximate scaling of~\cite{Girichidis20} to the loss rate solution of~\cite{Gould72} (see also~\citealp{Mannheim94}):
\begin{equation}
\label{eq:coulomb_lossrate}
    L_{\rm r,C}=19.7\,n_{\rm e}\left[1+\left(\frac{p}{{\rm GeV}/c}\right)^{-2}\right] \frac{\rm GeV}{c}\rm \,cm^3\,Gyr^{-1}\, ,
\end{equation}
where $n_{\rm e}$ is the free electron number density (see also 
\citealp{Winner19}).
Ionization losses follow a similar scaling with $L_{\rm r,ion}\simeq 0.57(n_{\rm neut}/n_{\rm e})L_{\rm r, C}$, where $n_{\rm neut}$ is the number density of neutrals, a contribution that we neglect in this paper assuming that gas is always fully ionized for clarity ($L_{\rm r,ion}=0$).

Hadronic losses correspond to the interaction of CR protons with thermalised particles, producing pions which decay into $\gamma$-ray photons.
The loss rate of CRs corresponding to hadronic interactions is:
\begin{equation}
\label{eq:hadronic_lossrate}
    L_{\rm r,h}= \frac{{\rm d}p}{{\rm d}T} c n_{\rm N}\sigma_{\rm pp}K_{\rm h}T(p)\Theta(p-p_{\rm thr})\,,
\end{equation}
where $\sigma_{\rm pp}\simeq 3.72\times 10^{-26}\,\rm cm^2$ is the total inelastic proton-proton cross-section, $K_{\rm h}=1/2$ is the inelasticity of the interaction, $\Theta$ is the heaviside function, $p_{\rm thr}=0.78 \,\rm GeV/c$ is the threshold momentum for hadronic losses, and $n_{\rm N}$ is the number density of nucleons~\citep{Mannheim94,Ensslin07,Guo08}.
We note that the hadronic interaction does not correspond in principle to a continuous shift in momentum since a CR proton loses half ($K_{\rm h}=1/2$) of its momentum in the interaction. 
However, since our momentum bins are sufficiently large in general (of the order of $\sim1\,\rm dex$), we can numerically treat this as a continuous process, and impose a $p$ bin-to-bin transfer of $f(p)$.

Finally, the full Coulomb losses and $1/6$-th of hadronic losses~\citep[e.g.][]{Kelner06} are added to the thermal energy ($\mathcal{H}_{\rm r}$).

\subsubsection{Spectrum time evolution}
\label{section:spectral_update}

The $\mathcal{Q}^n_{i\pm}$ and $\mathcal{Q}^e_{i\pm}$ terms (their $[...]_{\pim}^{\pip}$ component) lead to a time update $\Delta t_{\rm cr, spe}$ (we call it $\Delta t$ for simplicity here) at the boundaries of the momentum bin $p_i$, i.e.~$\nci(t+\Delta t)=\nci(t)+\Delta n_{{\rm c},{i+1/2}}-\Delta n_{{\rm c},{i-1/2}}$ and similarly for $\eci$, where 
\begin{equation}
    \Delta n_{{\rm c},{i-1/2}}=\int_t^{t+\Delta t}4\pi \pim^2 L(\pim) f(\pim){\rm d}t
\end{equation} 
and 
\begin{equation}
\Delta e_{{\rm c},{i-1/2}}=\int_t^{t+\Delta t}4\pi \pim^2 L(\pim) T(\pim) f(\pim){\rm d}t\, .
\end{equation}
Since $L(p)=-{\rm d}p/{\rm d}t$, we can use it to infer the relation between the time step and the momentum change with the following relation:
\begin{eqnarray}
    -\int_{p_{{\rm ini},i}}^{p_{{\rm fin},i}} \frac{{\rm d} p}{L(p)}=\int_t^{t+\Delta t} {\rm d}t =\Delta t\,,
\end{eqnarray}
where $p_{{\rm ini},i}$ and $p_{{\rm fin},i}$ are respectively the initial and final momentum values for a given bin $i$ over a time step $\Delta t$. 
Hence, the variations in $\nci$ and $\eci$ are now given by
\begin{equation}
\Delta n_{{\rm c},{i-1/2}}=\int_{\pim}^{p_{\rm ini}}4\pi p^2 f(p){\rm d}p    
\end{equation}
 and
\begin{equation}
\Delta e_{{\rm c},{i-1/2}}=\int_{\pim}^{p_{\rm ini}}4\pi p^2 T(p) f(p){\rm d}p    
\end{equation}
with $p_{\rm ini}=p_{\rm loss}>\pim$ for a loss term ($L(p)>0$), and $p_{\rm ini}=p_{\rm gain}<\pim$ for a gain term ($L(p)<0$).
Therefore, this requires to chose a time step to know the value of $p_{\rm ini}$ involved in the loss/gain process. 
However, the adopted time step must be chosen so that it does not overshoot the value of the momentum of the opposite boundary, i.e.~that $p_{\rm loss}<\pip$ for a loss process crossing the $i-1/2$ boundary of the bin, and with preferentially significantly smaller values. 
We opt for a value of $p_{\rm loss}<\pim+\varepsilon_p\Delta \log p_i$, and $p_{\rm gain}<\pim-\varepsilon_p\Delta \log p_i$, with $\varepsilon_p=0.1$, and we apply the smallest of the corresponding time step of each bin $p_i$ to the spectral update of all the $p_i$ bins.

The energy update by $\mathcal{Q}^e_\pm$ of a given bin $p_i$ has in addition the contribution from the integral term, which can be rewritten into:
\begin{equation}
    \frac{\partial \eci}{\partial t}=\frac{\Delta e_{{\rm c},{i+1/2}}}{\Delta t}-\frac{\Delta e_{{\rm c},{i-1/2}}}{\Delta t} + \mathcal{R}\eci
\end{equation}
where 
\begin{equation}
    \mathcal{R}=- \frac{\int^{\pip}_{\pim}p^{2}\beta(p)c L(p)f_0(p){\rm d}p}{\int^{\pip}_{\pim}p^{2} T(p)f_0(p){\rm d}p}\, .
\end{equation}
The results of these integrals are obtained by numerical integration (here with an adaptive Simpson's method to guarantee sub-percent accuracy).
The final value of $\eci(t+\Delta t)$ can be obtained with a semi-implicit time integration scheme, i.e.:
\begin{equation}
    \eci(t+\Delta t)=\frac{(1+0.5\mathcal{R}\Delta t)\eci(t)+\Delta e_{{\rm c},{i+1/2}}-\Delta e_{{\rm c},{i-1/2}}}{1-0.5\mathcal{R}\Delta t}\,.
\end{equation}
Alternatively, one could use an explicit or an implicit scheme, but we will show in Section~\ref{section:adiabatic} that they behave more poorly than our default choice.

Finally, we have to handle the boundary conditions in $p$-space, i.e.~the conditions for lowest ($i=1$) and highest ($i=N_{\rm c}$) bins. 
In practice we employ open boundary conditions in momentum space. 
We impose an identical slope to the lowest and highest boundary bins with their contiguous bins to compute the inflowing quantities when necessary, i.e.~for losses (radiative, streaming and adiabatic) cascading the momentum down at the highest bin boundary, and for gains (adiabatic) cascading the momentum up at the lowest bin boundary. 
Conversely, if the momentum flows down/up at the lowest/highest bin boundary, $n_{\rm c}$ and $e_{\rm c}$ are lost for the CR system.

\subsubsection{Spectral injection}

The injection of CRs is modeled by assuming a power-law momentum distribution of the form $j_0(p) = A_{\rm inj}\, p^{-q_{\rm inj}}$, which leads to analytical expressions for the injection rates of the CR number and energy densities, $j^n_{0,i} = \int_{\pim}^{\pip} 4\pi p^2 j_0(p)\, {\rm d}p$, $j^e_{0,i} = \int_{\pim}^{\pip} 4\pi p^2 T(p)\, j_0(p)\, {\rm d}p$.
The spectral slope $q_{\rm inj}$ is traditionally related to the shock compression ratio $r$ through  $q_{\rm inj} = 3r/(r - 1)$, yielding typical values of $q_{\rm inj} \simeq 4$ and $3.5$ for strong thermally- or CR-dominated shocks, respectively. 
In principle, the injection spectrum can therefore be determined self-consistently from the local shock properties  (e.g.~using a shock-finder algorithm as in~\citealp{Dubois19}). 
In the present work, however, we simply test our injection scheme with a constant value of $q_{\rm inj}$ and defer a self-consistent treatment to future work.

\subsubsection{Diffusion with the spectral method}
\label{section:diffusion}

The diffusion coefficients acting on $\nci$ and $\eci$ are equal when neglecting the inner slope (within the bin) of the distribution function and replacing it by a bin-centered Dirac delta function. 
This means that each $\kappa_{i}$ (also $\sigma_i$) is sampled at $p_i$ for commonly assumed power laws of the form $\kappa(p)=\kappa_0(p/p_0)^\delta$.  
A drawback of this bin-centered approximation is that the inner slope $q_i$ of the distribution function is conserved under diffusion, since both $\nci$ and $\eci$ experience the same diffusion coefficient (hence the ratio $\eci/\nci$ remains constant). 
In contrast, the global slope $\bar q_i = \log (f_{i+1}/f_{i-1})/\log (p_{i+1}/p_{i-1})$ is modified as an effect of the $p$-dependence of $\kappa$.  
Following the approach of~\citet{Hopkins23diffusion}, we include a more accurate treatment of diffusion by introducing their analytically derived correction terms (assuming all terms in the integral over momentum can be represented by a power-law) $\omega^n_i$ and $\omega^e_i$, leading to using modified diffusion coefficients $\tilde \kappa^n_i = \omega^n_i \kappa_i$, $\tilde \kappa^e_i = \omega^e_i \kappa_i$.
The correction factors $\omega^n_i$ and $\omega^e_i$ are of order unity and depend on the $p$-dependent slope of $\kappa_i$, on the slope of $f_0$ (i.e.~$q_i$), on the slope of the generating functions of $\nci$ and $\eci$ (i.e.~$1$ and $T(p)$, respectively), and on the bin size in $p$ \citep[see equation~31 in][]{Hopkins23diffusion}.

\section{Results}
\label{section:results}
\subsection{Idealised tests suite}
We start by testing the spectral CR implementation with a suite of idealised tests for cooling, injection, adiabatic changes, and diffusion.

\begin{figure}
    \centering
    \includegraphics[width=\columnwidth]{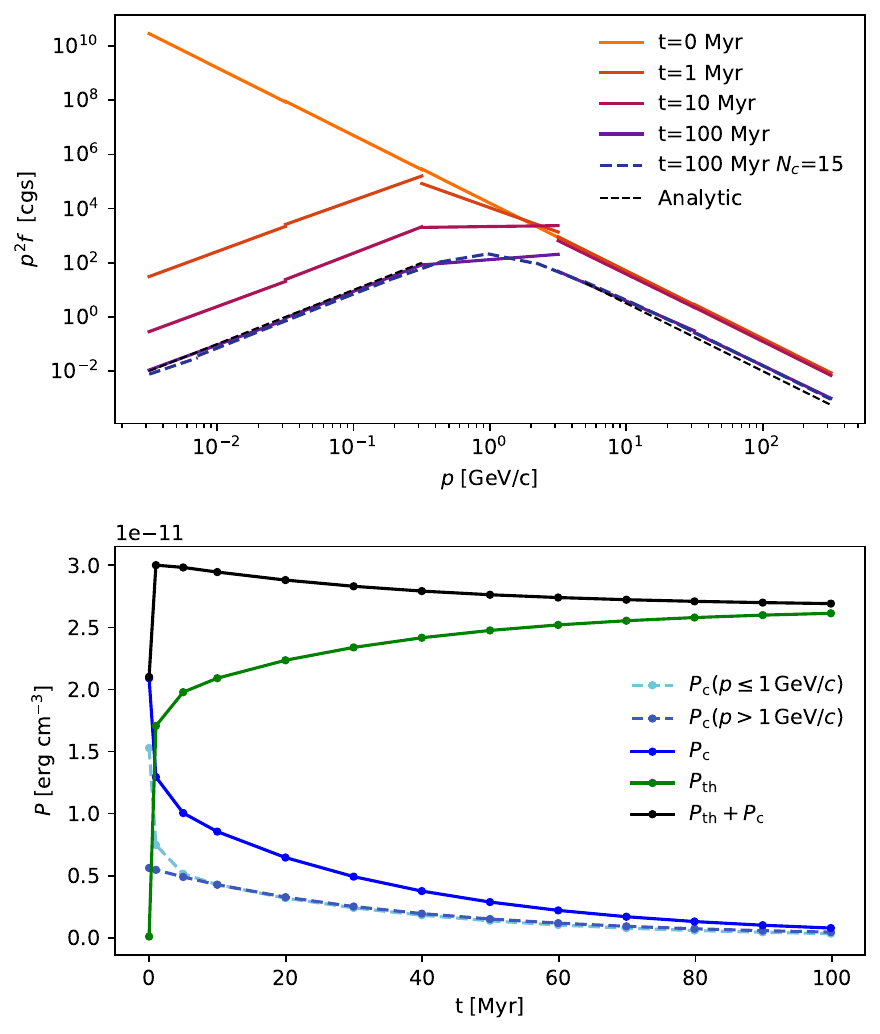}
    \caption{CR spectrum evolution cooled by Coulomb and hadronic losses, with $N_{\rm c}=5$ bins, an initial distribution function $f\propto p^{-4.5}$ and $n_{\rm e}=n_{\rm N}=1\,\rm cm^{-3}$. The top panel shows the evolution in time of the spectrum. The blue dashed curve shows the spectrum for $N_{\rm c}=15$ at $t=100 \, \rm Myr$ and the black dashed line correspond to the expected shape of the spectrum considering the radiative processes separately. We find that our test agrees well with the analytic prediction. In the bottom panel, we show the time evolution of the total CR pressure, the pressures of the low-momentum and high-momentum CR components, and the gas pressure, which illustrate the transfer of energy from CRs to the thermal plasma.}
    \label{fig:freecool}
\end{figure}

\subsubsection{Radiative losses}
\label{section:freecooling}

We test the free cooling of CR protons and of their distribution function by Coulomb and hadronic radiative losses (we also show in Appendix~\ref{appendix:cr_electrons} the same test for the evolution of CR electrons under synchrotron losses).
In practice, this test solves the following reduced equations of CR evolution:
\begin{align}
&\frac{\partial \nci}{\partial t} =\left[4\pi p^2 L_{\rm r} f_0\right]_{\pim}^{\pip}\, , \nonumber\\
&\frac{\partial \eci}{\partial t} =\left[4\pi p^2 L_{\rm r} T(p) f_0\right]_{\pim}^{\pip}- 4\pi \int^{\pip}_{\pim}p^2 v L_{\rm r} f_0 {\rm d}p\, . \nonumber
\end{align}
The distribution function starts with an initial continuous power-law index of $q_{\rm ini}=4.5$ ranging from $p_{\rm min}=10^{-2}\,\rm GeV/c$ and $p_{\rm max}=10^{2}\,\rm GeV/c$ with $N_{\rm c}=5$ the number of bins, and the gas number densities for nucleons and free electrons are identical, i.e.~assuming that the gas is static with $n_{\rm e}=n_{\rm N}=1 \,\rm cm^{-3}$.

In the top panel of Fig.~\ref{fig:freecool}, we show the evolution of the CR spectrum at several times up to $t=100\,\rm Myr$.
For comparison, we also plot the CR spectrum at the final time obtained with $N_{\rm c}=15$, using a $p$-binning such that the two extreme momentum boundaries match those of the $N_{\rm c}=5$ case.
The black dashed lines indicate the expected slopes of the cooled spectrum (see Appendix~\ref{appendix:analytical_solution}).
Quickly, the numerical solution for the distribution function $f(p)$ exhibits a constant value at low energies, where Coulomb losses dominate ($p\lesssim1\,\rm GeV/c$), while in the ultra-relativistic regime, dominated by hadronic losses, the spectrum preserves its initial slope with a reduced normalization.
Both regimes are in very good agreement with the analytical expectations.
Remarkably, this analytical behaviour is already well captured with very coarse momentum binning, with bins spanning one dex in size ($N_{\rm c}=5$).
The intermediate slopes smoothly connect to the large-scale slope of the distribution function, with the bending momenta occurring at $p\simeq0.28$, $0.66$, and $2.4\,\rm GeV/c$ at $t=1$, $10$, and $100\,\rm Myr$, respectively, in agreement with Eq.~\eqref{eq:coulomb_lossrate}.

In the bottom panel of Fig.~\ref{fig:freecool}, we present the time evolution of the total CR pressure and the gas pressure.
Radiative losses transfer CR energy and pressure to the thermal component.
Total energy is not conserved, since only one sixth of the hadronic energy losses are returned to the gas\footnote{We verified that exact energy conservation is recovered if all hadronic losses are deposited into the thermal component.}.
We also show the time evolution of the CR pressure for both low-momentum and high-momentum CRs, which highlights that the energy losses primarily arise from low-momentum CRs due to Coulomb interactions and occur more abruptly, whereas the energy losses of high-momentum CRs are smoother.
Because the adiabatic index of the gas is larger than that of the CRs, the total pressure remains closer to constant than the total energy.
Nevertheless, the detailed evolution and balance between the two components depend on the initial shape of the CR distribution function.

\begin{figure}
    \centering
    \includegraphics[width=\columnwidth]{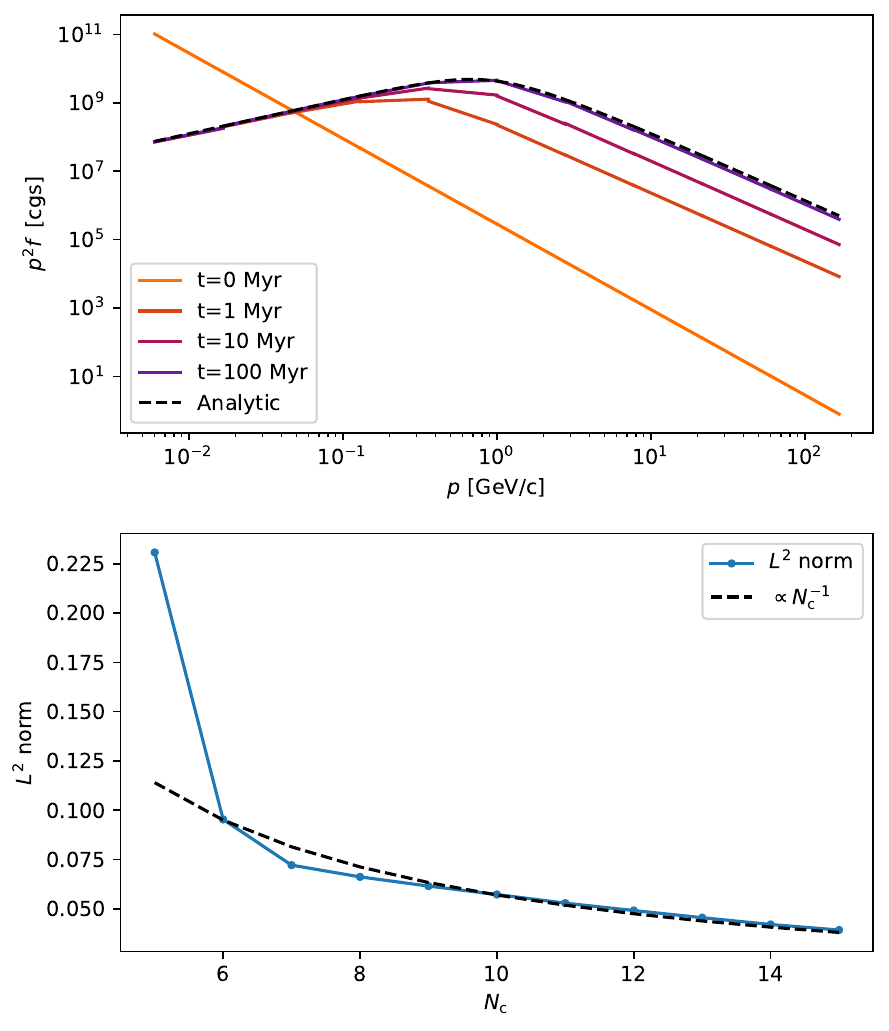}
    \caption{CR spectrum evolution in case of injection and radiative losses, with  $N_{\rm c}=10$ bins and $n_{\rm e}=n_{\rm N}=1\,\rm cm^{-3}$. The first panel shows the time evolution of the spectrum with an injection slope $q_{\rm inj}=4$. The black dashed line correspond to the analytic solution of equation~\eqref{eq:steadystate_solution}. The second panel shows the error of the numerical solution compared to the analytic result using the $L^2$ norm.}
    \label{fig:steadystate_q4}
\end{figure}

\subsubsection{Steady-state solution with injection and cooling}
\label{section:steady}

We consider a spatially homogeneous system subject to continuous CR injection and radiative energy losses, in the absence of transport or escape processes.
This test solves the following reduced equations of CR evolution:
\begin{align}
&\frac{\partial \nci}{\partial t} =\left[4\pi p^2 L_{\rm r} f_0\right]_{\pim}^{\pip}+j^{\rm n}_{0,i}\, , \nonumber\\
&\frac{\partial \eci}{\partial t} =\left[4\pi p^2 L_{\rm r} T(p) f_0\right]_{\pim}^{\pip}- 4\pi \int^{\pip}_{\pim}p^2 v L_{\rm r} f_0 {\rm d}p+j^{\rm e}_{0,i}\, . \nonumber
\end{align}
The injection rate is assumed to be time-independent and to follow a power law in momentum, $j_0(p) = A p^{-q_{\rm inj}}$, while energy losses are described by a momentum-dependent radiative loss term $L_{\rm r}(p)$.
Under these conditions, the CR distribution function evolves solely through the competition between injection and radiative cooling.

At steady state, this balance leads to a momentum-dependent equilibrium spectrum whose shape is controlled by the dominant loss process, and writes:
\begin{equation}
\label{eq:steadystate_solution}
f_{\rm eq}(p)=\frac{A\, p^{1-q_{\rm inj}}}{(q_{\rm inj}-3)\,L_{\rm r}(p)} \, .
\end{equation}
In the sub-relativistic regime, Coulomb losses dominate and scale as $L_{\rm r} \simeq L_{\rm r,C} \propto p^{-2}$, resulting in a flattened steady-state spectrum with slope $3 - q_{\rm inj}$.
In contrast, at relativistic momenta where hadronic losses dominate and scale as $L_{\rm r} \simeq L_{\rm r,h} \propto p$, the steady-state spectrum retains the original injection slope $-q_{\rm inj}$.
This behaviour reflects the strong momentum dependence of Coulomb losses at low energies, which efficiently redistribute particles toward higher momenta, whereas hadronic losses act proportionally to momentum and therefore preserve the injected spectral shape.

In Fig.~\ref{fig:steadystate_q4}, we show the evolution of the CR distribution function under continuous injection and cooling for $q_{\rm inj}=4$, $A=10^{35}(m_{\rm H}c)^{q_{\rm inj}}\,\rm Myr^{-1}$, and $n_{\rm e}=n_{\rm N}=1\,\rm cm^{-3}$, together with the corresponding analytical steady-state solution. 
The top panel displays the CR distribution function at different times, illustrating the progressive convergence toward the steady-state solution as radiative losses reshape the injected spectrum.
By $t=100\,\rm Myr$, the numerical solution has reached steady state and is almost indistinguishable from the analytical prediction over the full momentum range.
This timescale is set by the longest radiative cooling time in the system, which corresponds to the hadronic loss timescale for the chosen physical parameters and is $t_{\rm h}\simeq57\,\rm Myr$.

To provide a more quantitative assessment of the numerical accuracy, the bottom panel of Fig.~\ref{fig:steadystate_q4} shows the $L^2$ norm of the difference between the numerical and analytical distribution functions for different numbers of momentum bins.
The $L^2$ norm is computed over the full momentum range as the quadratic norm of the difference between the two solutions, normalized by the analytical steady-state solution.
We find that the error decreases monotonically with increasing resolution and scales approximately as $\propto N_{\rm c}^{-1}$.
This behaviour indicates a linear convergence of the solution error with $\Delta\log p$ over the explored range of resolutions.

\subsubsection{Adiabatic changes and time integration scheme}
\label{section:adiabatic}

Adiabatic changes (and similarly, hadronic losses) in the CR distribution function correspond to reversible compression or expansion of the gas, which acts as an effective advection process in momentum space. Specifically, during compression, CRs are shifted toward higher momenta, increasing the amplitude of the distribution function $f(p)$, whereas during expansion, they are shifted toward lower momenta, decreasing\footnote{This holds for sufficiently steep distributions with $q>3$; otherwise, compression can decrease the amplitude of the distribution function and conversely for expansion (see Appendix~\ref{appendix:freecooling_slope}).} $f(p)$.
Because adiabatic processes conserve phase-space density, these transformations are exactly reversible in the absence of additional energy-changing processes, and the shape of the distribution function is preserved (see Appendix~\ref{appendix:analytical_solution}).

We now solve the reduced equations governing CR evolution:
\begin{align}
&\frac{\partial \nci}{\partial t} =\left[4\pi p^3\frac{\vec\nabla.\vec u}{3} \ f_0\right]_{\pim}^{\pip}\, , \nonumber\\
&\frac{\partial \eci}{\partial t} =\left[4\pi p^3\frac{\vec\nabla.\vec u}{3} T(p) f_0\right]_{\pim}^{\pip}- 4\pi \int^{\pip}_{\pim}p^3v\frac{\vec\nabla.\vec u}{3} f_0 {\rm d}p\, . \nonumber
\end{align}
To test the numerical implementation, its scaling with the time step and the impact of the time integration scheme, we impose a compression regime with a constant velocity divergence $\vec \nabla . \vec u = -1\,\rm Myr^{-1}$. The choice of adiabatic compression for this test is arbitrary; an equivalent setup could have been realized with expansion.
The initial CR distribution, $f \propto p^{-4.5}$, is discretized into $N_{\rm c}=15$ bins spanning $p_{\rm min} = 10^{-2}\,{\rm GeV}/c$ to $p_{\rm max} = 10^{2}\,{\rm GeV}/c$. The gas is kept static, so no hydrodynamic feedback or additional CR processes (e.g.~cooling or injection) are included.

\begin{figure}
    \centering
    \includegraphics[width=\columnwidth]{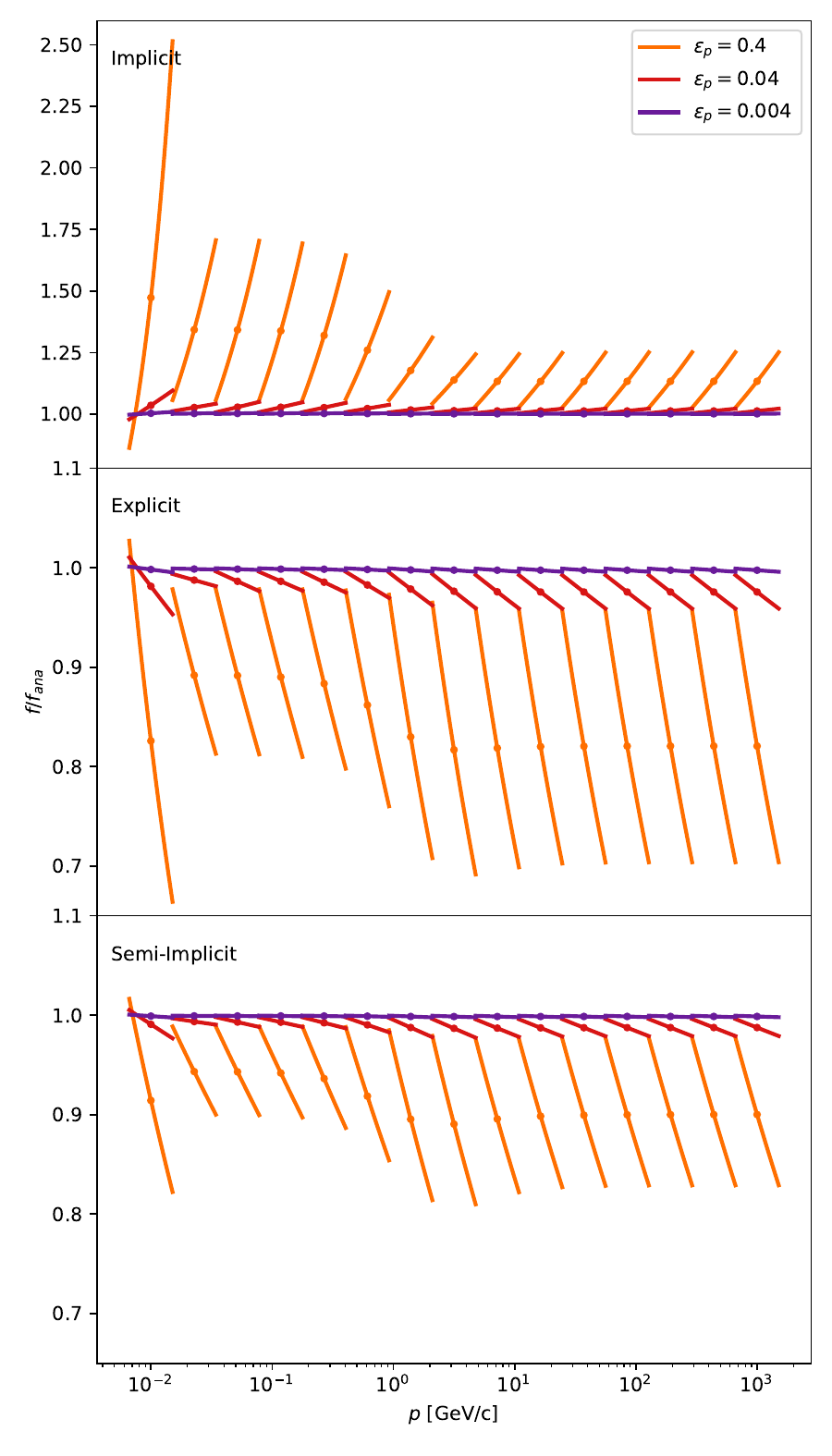}
    \caption{Static test with adiabatic compression cycles using $\vec \nabla.\vec u =-1\,\rm Myr^{-1}$. The panels show the relative ratio of the distribution function at a given time with respect to the analytical distribution function $f_{\rm ana}$ for $q_{\rm ini}=4.5$. The dots indicate the value of the ratio at the center of the bin. We present the solution for three spectral time steps and three time integration schemes.}
    \label{fig:periodic_adia}
\end{figure}

Figure~\ref{fig:periodic_adia} shows the evolution of the distribution function up to $1\,\rm Myr$, plotted as the ratio of $f(p,t)$ to the analytical solution (see Appendix~\ref{appendix:hadronic_prediction}). 
We present this ratio for different values of $\varepsilon_p$ to assess the effect of the spectral time step and to compare various time integration schemes for updating $\eci$.
A systematic offset is observed in the ratio, scaling linearly with the spectral time step. For a given integration scheme, the smallest offset is obtained for $\varepsilon_p = 0.004$.
The choice of time integration scheme also significantly affects the results. For larger $\varepsilon_p$ (i.e.~larger time steps), the implicit method produces substantial errors, particularly at low momenta.
Using an explicit method considerably reduces these errors, while a semi-implicit method further decreases them to about $10\%$. For smaller time steps, the errors fall below $1\%$ with the semi-implicit method.

\subsubsection{One-dimensional diffusion}
\label{section:1d_diffusion}

\begin{figure}
    \centering
    \includegraphics[width=\columnwidth]{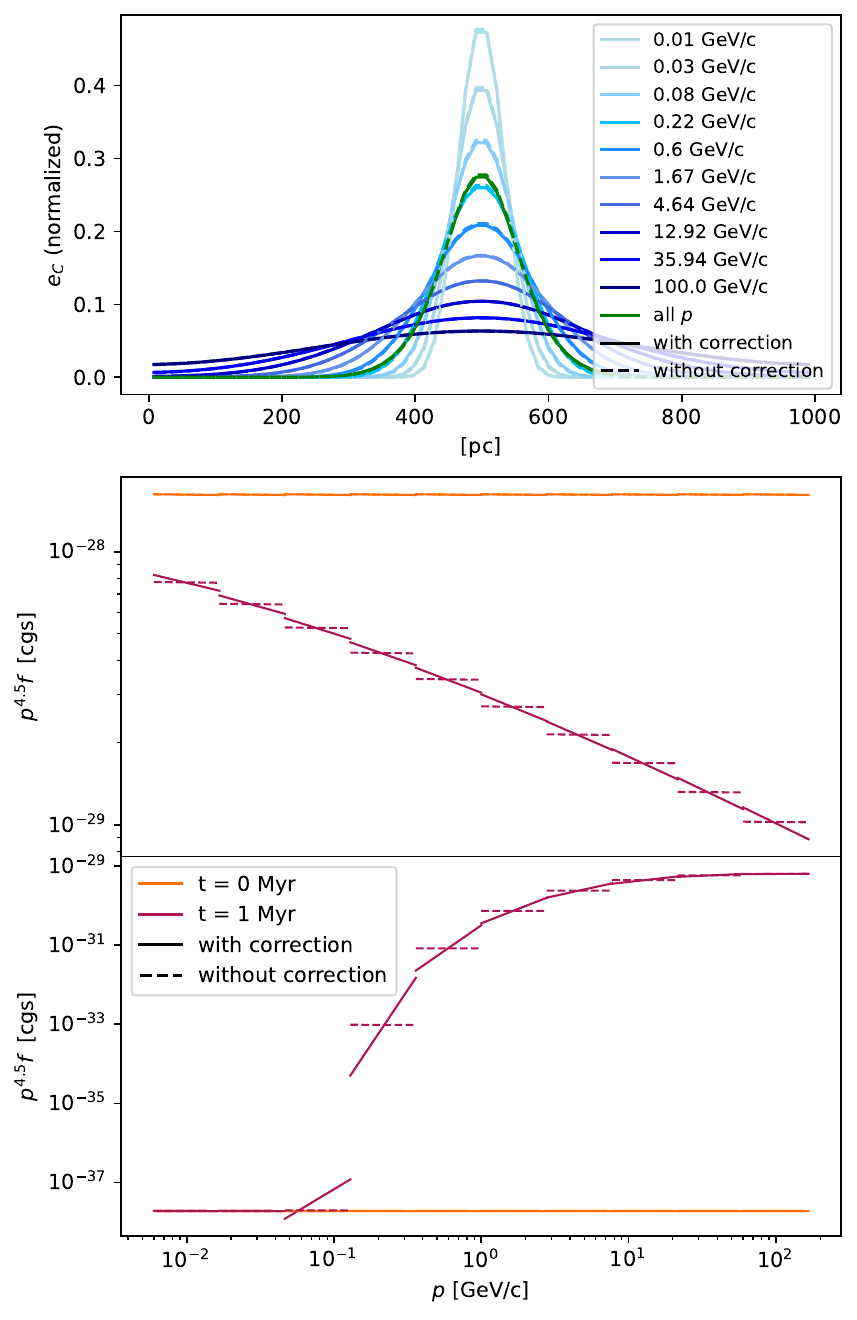}
    \caption{Spatial diffusion in one-dimension of the CR energy densities $\eci$ at $t=1\,\rm Myr$ for each momentum bin, assuming that diffusion scales as $\kappa(p)\propto p^{0.5}$.
    The green curves show the total CR energy density (sum over all bins). All CR quantities are normalized to their initial values at the peak of the gaussian. The two lower panels show the CR distribution function at two different positions $x=500\,\rm pc$ and $x=750\,\rm pc$ (middle and bottom panels respectively). Solid lines correspond to the spectrally-averaged diffusion coefficients (``with correction''), while dashed lines assume $\kappa^\nci=\kappa^\eci$ (``without correction'').
}
    \label{fig:diff_ecr_1d_spectra}
\end{figure}

We performed a one-dimensional diffusion simulation to illustrate the effect of the multi-group spectral discretization on the spatial transport of CRs.
This problem solves for this set of two-moment transport equations for $\nci$ and $\eci$ and their associated fluxes:
\begin{align}
&    \frac{\partial \nci}{\partial t} +\frac{\partial F^\nci}{\partial x}=0\, ,\nonumber\\
&    \frac{1}{v^2}\frac{\partial F^\nci}{\partial t} +\frac{1}{3}\frac{\partial \nci}{\partial x} = -\frac{1}{3\kappa^\nci} F^\nci \, ,\nonumber\\
&    \frac{\partial \eci}{\partial t} +\frac{\partial F^\eci}{\partial x}= 0\, , \nonumber\\
&    \frac{1}{v^2}\frac{\partial F^\eci}{\partial t} +\frac{1}{3}\frac{\partial \eci}{\partial x}= -\frac{1}{3\kappa^\eci} F^\eci\, .\nonumber
\end{align}
The gas was set up static and no cooling or additional CR processes were included.
The spatial domain was covering $1\,\rm kpc$ with $64$ spatial cells and outflow boundary conditions were used.
The CRs were initialized with a power-law distribution function of index $q_{\rm ini}=4.5$ in momentum space, spanning $p_{\rm min}=0.01\,\rm GeV/c$ to $p_{\rm max}=100\,\rm GeV/c$ with $N_{\rm c}=10$ bins.
Spatially, the CRs were initially distributed as a gaussian of width $\sigma=10\,\rm pc$ with a total energy at the peak of the gaussian $e_{\rm c,max}=2.5 \times 10^{-10}\, \rm erg\,cm^{-3}$ at the center (but note that the problem is independent of the exact normalization).
We adopted a reduced speed of light $\tilde c = 3\times10^3\,\rm km\,s^{-1}$, and the diffusion coefficient was specified as $\kappa(p)=\kappa_0(p/p_0)^{0.5}$, with $\kappa_0=10^{27}\,\rm cm^2\,s^{-1}$ at $p_0=1\,\rm GeV/c$.

Considering that spectrally resolved CRs affect spatial diffusion, as explained in Sec.~\ref{section:diffusion}, we now distinguish between the diffusion coefficients for the number density and the energy density, $\kappa^\nci \neq \kappa^\eci$.
These coefficients are obtained through different spectral weightings of $\kappa(p)$, following the procedure of~\cite{Hopkins23diffusion}, which is our default approach.
We compare this approach to the simpler assumption $\kappa^\nci = \kappa^\eci = \kappa(p_i)$ at the center of each momentum bin.

Figure~\ref{fig:diff_ecr_1d_spectra} (top panel) shows the final spatial distribution of the energy density $\eci(x)$ at $t=1\,\rm Myr$ for each momentum bin, along with the total CR energy density (green curves).
The solid lines correspond to the spectrally-averaged diffusion coefficients, while the dashed lines correspond to $\kappa^\nci=\kappa^\eci$.
Fitting the total CR energy profile with a gaussian yields an effective diffusion coefficient $\kappa_{\rm eff} \simeq 5.4\times 10^{26}\,\rm cm^2\,s^{-1}$.
We find that using spectrally-averaged coefficients does not lead to a large difference in the spatial distribution of $\eci(x)$ compared to the simplified approach, confirming that the main effect of spectral averaging manifests in the shape of the distribution function rather than in the overall spreading of CR energy.
As expected, the highest momentum bins, which have larger diffusion coefficients, diffuse more rapidly and are more spatially extended than the lowest momentum bins, demonstrating that the spectral dependence of $\kappa(p)$ produces, indeed, a momentum-dependent diffusion signature.

To quantify how the spectral resolution affects the distribution function locally, we consider two representative positions: the center of the gaussian ($x=500\,\rm pc$) and the wings ($x=750\,\rm pc$).
The middle and bottom panels of Fig.~\ref{fig:diff_ecr_1d_spectra} show the CR distribution function at these positions, respectively, both initially and at $t=1\,\rm Myr$.
While both approaches reproduce the global slope of the spectrum reasonably well, differences appear in the inner slope of $f(p)$.
Using the spectrally-averaged diffusion coefficients correctly captures the momentum-dependent spreading, allowing the inner slope to reflect the global distortion induced by diffusion.
In contrast, assuming $\kappa^\nci=\kappa^\eci$ preserves the initial slope $q_{\rm ini}$ locally, failing to account for the differential diffusion between bins.
This demonstrates that spectral averaging of the diffusion coefficients is important for accurately capturing the shape evolution of the CR distribution function, even when the overall energy density profile is only mildly affected.

\begin{figure*}
\centering \includegraphics[width=2.0\columnwidth]{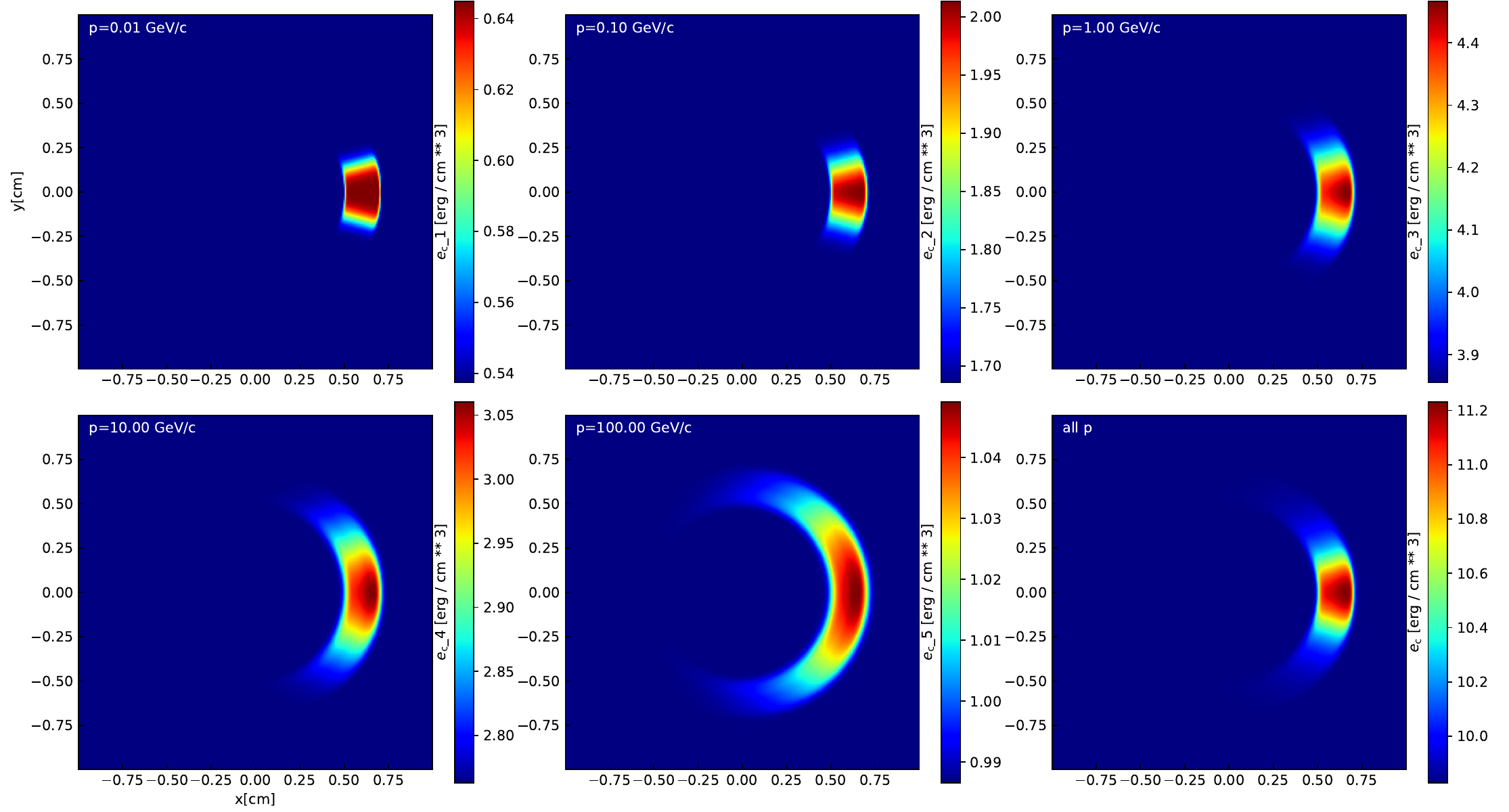}
    \caption{Spatial distribution of CR energy density for the two-dimensional anisotropic diffusion in a circular magnetic field. The five first panels show the results for a different momentum bin and last one is the CR energy density in total, all at the same time $t=0.26 \,\rm s$.}
    \label{fig:diff_donut_correct}
\end{figure*}

\subsubsection{Two-dimensional diffusion with anisotropic diffusion}
\label{section:2d_diffusion}

We now test the multi-group implementation of anisotropic diffusion by simulating the evolution of a patch of CRs in a circular magnetic field in two dimensions, following the setup of~\Rosdahl.
This two-dimensional problem solves for this set of two-moment transport equations for $\nci$ and $\eci$ and their associated fluxes:
\begin{align}
&    \frac{\partial \nci}{\partial t} +\vec \nabla . (F^\nci\vec b)=0\, ,\nonumber\\
&    \frac{1}{v^2}\frac{\partial F^\nci}{\partial t} +\vec b.\vec \nabla \left(\frac{\nci}{3}\right) = -\frac{1}{3\kappa^\nci} F^\nci \, ,\nonumber\\
&    \frac{\partial \eci}{\partial t} +\vec \nabla.(F^\eci\vec b)= 0 \, , \nonumber\\
&    \frac{1}{v^2}\frac{\partial F^\eci}{\partial t} +\vec b .\vec \nabla \left (\frac{\eci}{3} \right)= -\frac{1}{3\kappa^\eci}F^\eci\, . \nonumber
\end{align}

The computational domain was set up as a square box of width $2\,\rm cm$ with a circular magnetic field centered in the domain.
We used a spatial resolution of $512\times 512$ cells.
The gas was set up homogeneous and static, and no cooling is included, so the evolution is purely diffusive.
The CR energy density is initialized with a high-energy patch of $e_{\rm c,in}=12\,\rm erg\,\rm cm^{-3}$ in the region defined by $0.5 < r < 0.7\,\rm cm$ and $\vert\theta\vert<\pi/12$ (angle with respect to the $x$-axis), while the surrounding medium has $e_{\rm c,out}=10\,\rm erg\,\rm cm^{-3}$.
The CRs are discretized into $N_{\rm c}=5$ momentum bins between $p_{\rm min} = 10\,{\rm MeV}/c$ and $p_{\rm max} = 100\,{\rm GeV}/c$, with an initial power-law slope $q_{\rm ini}=4.5$.
The diffusion coefficient is specified as $\kappa=3.33\times 10^{-2}(p/p_0)^{0.5}\,\rm cm^2\,s^{-1}$ with $p_0=1\,{\rm GeV}/c$, and the simulation uses a reduced speed of light $\tilde c = 100\,\rm cm\,s^{-1}$.
In this setup, the spectral correction for the diffusion coefficients is applied, meaning that the effective diffusion coefficient for each bin is computed using the bin-averaged approach as in Sec.~\ref{section:diffusion}.

Figure~\ref{fig:diff_donut_correct} shows the CR energy distribution at $t=0.26\,\rm s$.
The first five panels display the energy density for each momentum bin, while the last panel shows the total CR energy.
As expected, higher-momentum bins diffuse faster and spread further along the circular magnetic field lines, reflecting the momentum dependence of $\kappa(p)$.
The total CR energy density maintains the overall shape of the initial patch while exhibiting a smoothed profile, demonstrating that the spectral method captures both anisotropic and momentum-dependent diffusion accurately.

This test confirms that the scheme correctly models the expected anisotropic spreading of CRs along magnetic field lines and reproduces the dependence of diffusion on CR momentum, validating the multi-group treatment of the diffusion of CRs in multidimensional configurations.

\begin{figure}
    \centering
    \includegraphics[width=0.49\columnwidth]{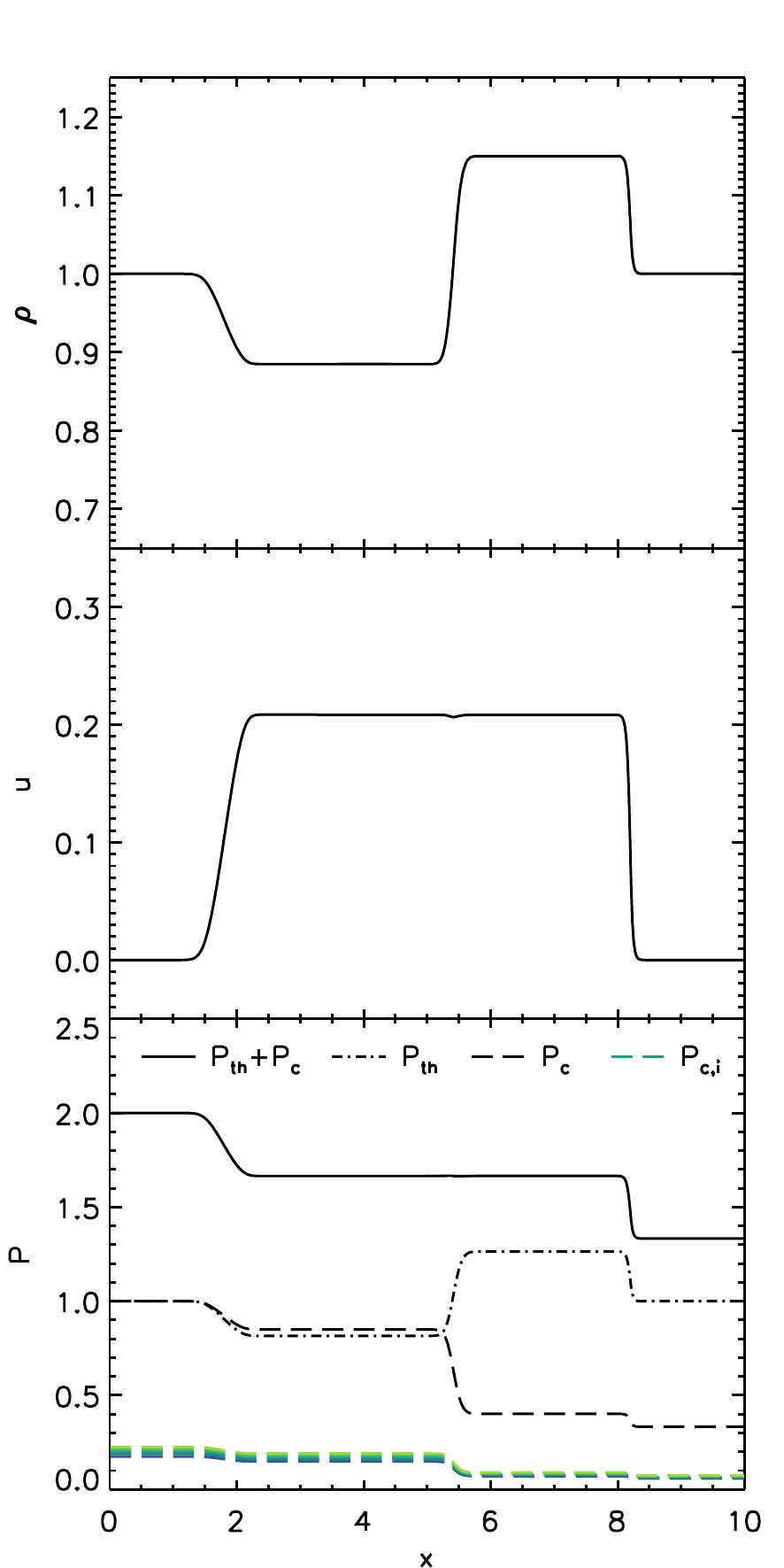}
    \includegraphics[width=0.49\columnwidth]{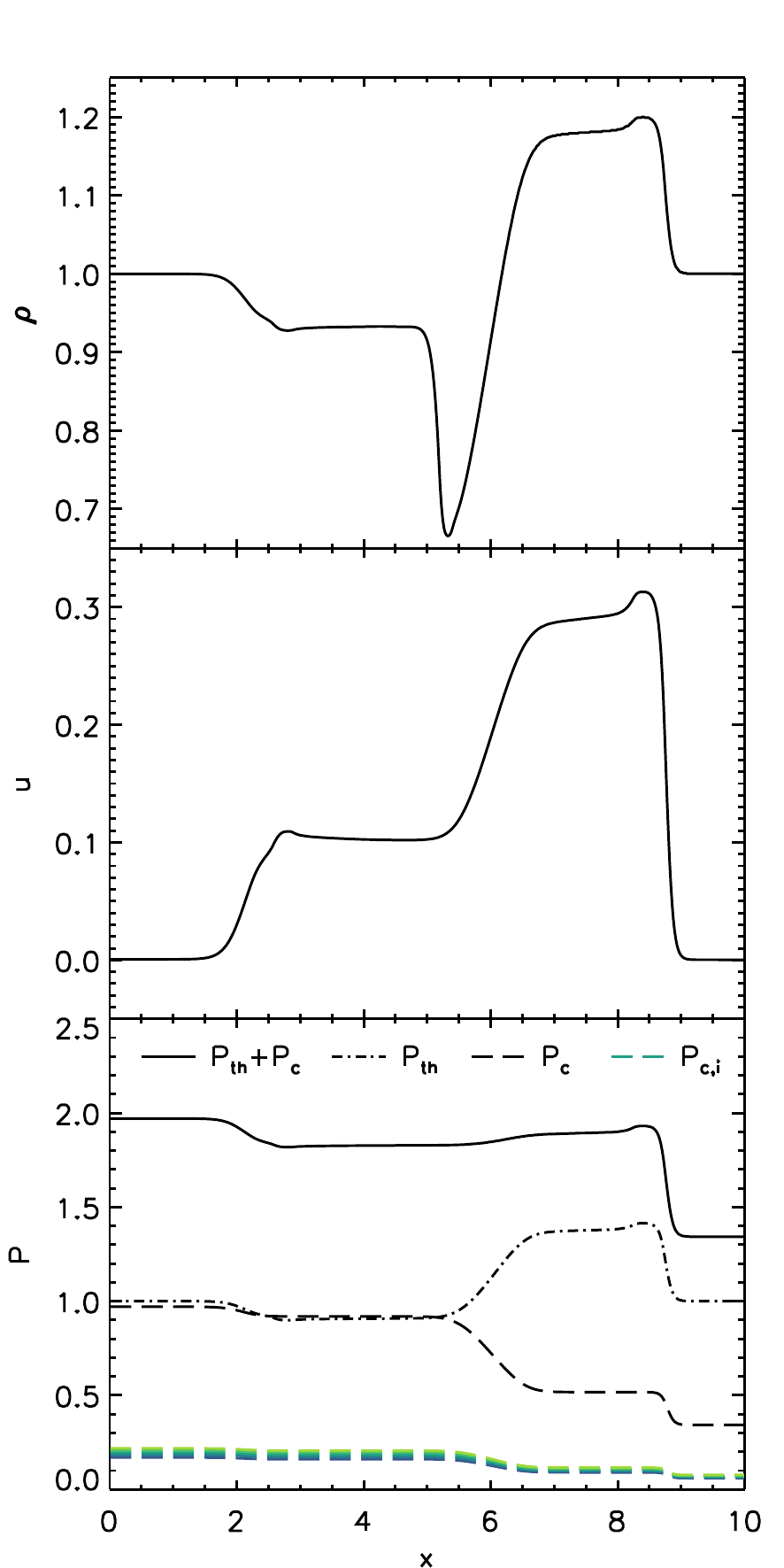}
    \caption{Shock tube test run with the spectral method. We follow $N_{\rm c}=5$ momentum bins initialised with $f \propto p^{-4}$ and show the result for $t=2 \rm\, s$. The panels show from top to bottom gas density, gas velocity, the total pressure (solid), for the thermal component (dot-dashed), and for CRs (dashed) with each pressure of individual CR group represented in blue-green shades from the lowest to the highest $p$-bins (we have slightly over-set them but they nearly perfectly overlap in practice). The first column is for the simulation with constant Alv\'en velocity of 0 (no streaming), while the second column is with $u_{\rm A}=0.75\,\rm km\,s^{-1}$ (with streaming). }
    \label{fig:shock_tube}
\end{figure}

\subsubsection{Shock tube with streaming}
\label{section:shock_tube}

We consider a standard one-dimensional shock tube test similar to that of~\cite{Thomas21}, which was also performed using the grey CR implementation in {\sc ramses} in~\Rosdahl, and extend it here to the spectral multi-group CR formulation. 
We solve here the full CRMHD equations without including radiative losses ($L_{\rm r}=0$).
The initial state consisted of a homogeneous, static gas with density $\rho = 1\,\rm g\,cm^{-3}$, thermal pressure $P_{\rm th}=1\,\rm erg\,cm^{-3}$, and zero velocity.
A shock was generated by imposing an initial discontinuity in the total CR pressure: the left half of the domain is initialized with $P_{\rm c,tot}=1\,\rm erg\,cm^{-3}$, while the right half has $P_{\rm c,tot}=1/3\,\rm erg\,cm^{-3}$.

To closely match the setup of~\Rosdahl, the CR population is chosen to be fully ultra-relativistic, so that the CR adiabatic index is $\gamma_{\rm c}\simeq4/3$.
This requires both a sufficiently large momentum range and a distribution function with slope $q=4$.
This choice is also motivated by the form of the streaming advection speed in {\sc ramses-mcr}, which scales as $q\,u_{\rm A}/3$, whereas in grey formulations it is commonly written as $\gamma_{\rm c} u_{\rm A}$.
For $q=4$, these two expressions coincide.
We therefore discretize the CR distribution into $N_{\rm c}=5$ momentum bins spanning $p_{\rm min}=100\,\rm GeV/c$ to $p_{\rm max}=100\,\rm TeV/c$, ensuring that all bins remain in the ultra-relativistic regime.
The initial distribution function is $f \propto p^{-4}$, and we adopt a momentum-independent diffusion coefficient $\kappa(p)=\kappa_0$ with $\kappa_0=3.33\times10^{-6}\,\rm cm^2\,s^{-1}$.
The reduced speed of light is set to $\tilde c=100\,\rm cm\,s^{-1}$.
The spatial domain extends over $10\,\rm cm$ and is resolved with 512 uniform cells.
We perform three simulations with constant Alfvèn speeds $u_{\rm A}=0$ (no streaming) and $u_{\rm A}=0.75\,\rm km\,s^{-1}$ (we also tested $u_{\rm A}=1.5\,\rm km\,s^{-1}$ and verified that the results, not shown here, are consistent with those of~\Rosdahl).

The resulting gas density, velocity, and pressure profiles at $t=2\,\rm s$ are shown in Fig.~\ref{fig:shock_tube}.
In the absence of streaming (left panels), the solution exhibits the expected wave pattern: a shock propagating to the right ($x\simeq8\,\rm cm$), a contact discontinuity located near $x\simeq5.5\,\rm cm$, and a rarefaction wave moving to the left ($x\simeq2\,\rm cm$).
Because the distribution function has slope $q=4$ and the momentum bins are evenly spaced in $\log p$, each bin contributes equally to the total CR pressure, so that the individual CR pressures $P_{{\rm c},i}$ are identical.

When CR streaming is enabled (right panels), the solution is qualitatively modified.
As previously observed in~\Rosdahl using the grey method, a depletion in gas density and a jump in velocity develop near the location of the contact discontinuity.
This feature arises from the transfer of CR energy to the thermal gas through Alfvèn wave damping (Eq.~\ref{eq:stream_loss}).
Since the thermal adiabatic index satisfies $\gamma > \gamma_{\rm c}$, this energy transfer locally steepens the total pressure gradient, producing an enhanced acceleration of the gas away from the discontinuity.
In addition, the positions of both the shock and the rarefaction wave are shifted relative to the non-streaming case, reflecting the additional CR transport induced by streaming at the Alfv\'en speed.
Not shown in Fig.~\ref{fig:shock_tube} is that the slope of the CR distribution function is preserved throughout the evolution, indicating that streaming acts as a pure transport process.
Additional tests exploring different slopes of the distribution function and extended momentum ranges, including sub-relativistic CRs, are presented in Appendix~\ref{appendix:shock_streaming_variants}.
Overall, this test demonstrates that the multi-group CR implementation accurately reproduces the known behaviour of CR-modified shock tubes with streaming, while consistently accounting for the momentum-space structure of the CR population.

\begin{figure*}
    \centering
    \includegraphics[width=2\columnwidth]{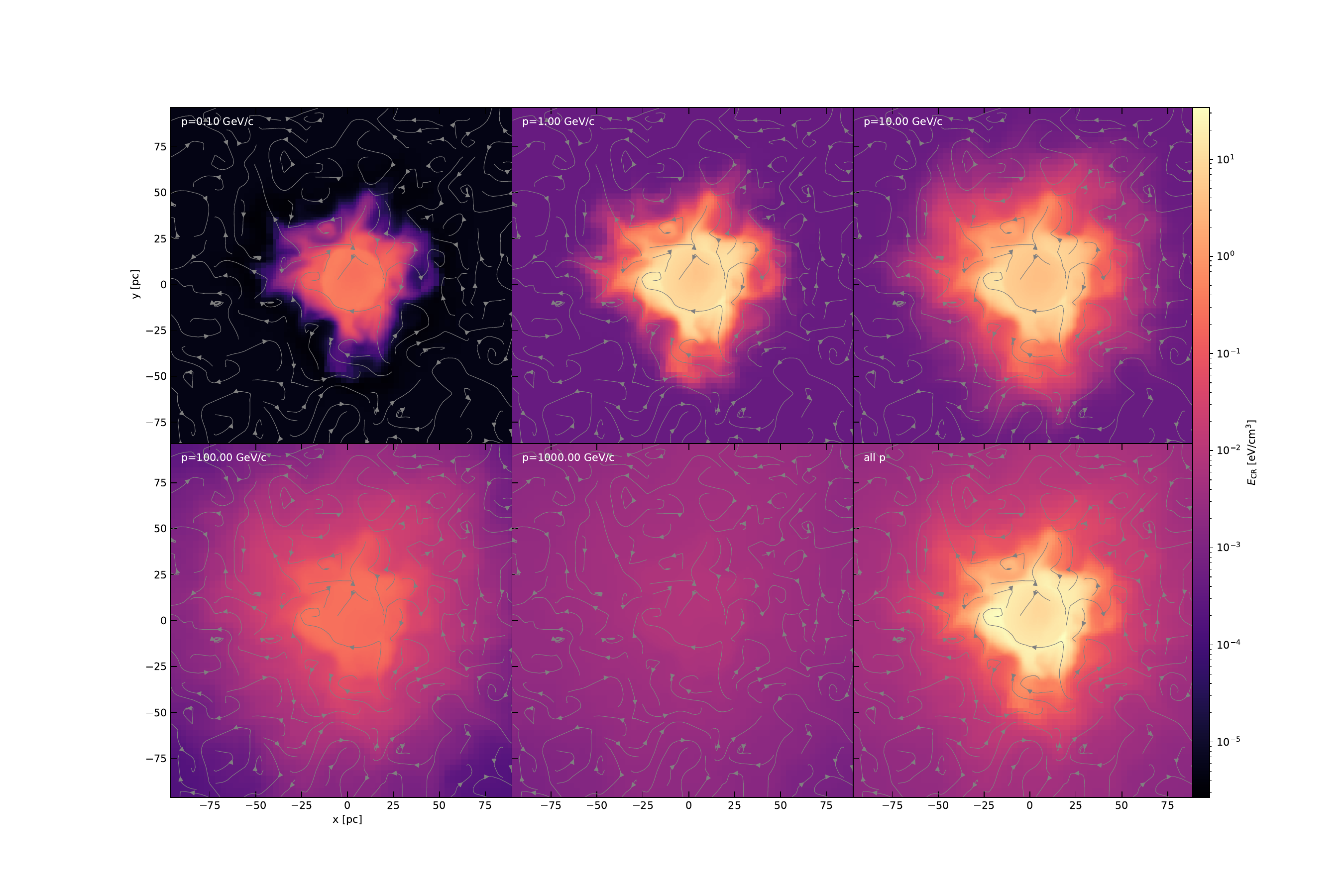}
    \caption{Three-dimensional SN explosion with CR and gas cooling at $t=200\,\rm kyr$ for gas density $n=10\,\rm cm^{-3}$ and a box size of $200\,\rm pc$ with random magnetic field (represented by grey streamlines). We use the spectral method with a scaling of diffusion of $\kappa(p)=3\times 10^{26}(pc/1 {\rm GeV})^{0.5}\,\rm cm^2\,s^{-1}$. We follow 5 bins in $p$ with $p_{\rm min}=100\,{\rm MeV}/c$ and $p_{\rm max}=1\,{\rm TeV}/c$: from top left to bottom middle corresponding to the CR energy density of each CR energy-momentum bin in a slice through the center of the explosion. Bottom right panel is the total CR energy density.}
    \label{fig:sedD26}
\end{figure*}

\subsection{A supernova remnant with cosmic rays}

We now use the full spectral multi-group CR formalism coupled to the MHD solver to follow the expansion of a SNR including CRs, their anisotropic diffusion, and their energy losses.
CR streaming is not included in this experiment, so that CR transport is entirely governed by advection with the gas and spatial diffusion.
This setup provides a controlled yet astrophysically relevant demonstration of the capabilities of {\sc ramses-mcr}.

The three-dimensional SNR experiment is performed in a uniform background medium with gas number density $n_0=10\,\rm cm^{-3}$ and temperature $T_0=100\,\rm K$, within a cubic box of side length $200\,\rm pc$.
Background CRs are initialized with a total CR energy density of $e_{{\rm c},0}=0.01\,\rm eV\,cm^{-3}$ and a power-law momentum distribution $f(p)\propto p^{-q_{\rm ini}}$ with slope $q_{\rm ini}=4.5$.
This value of background CRs is chosen artificially low so that the injected SN CRs, when spread by diffusion, dominate over the initial value of the background CRs. 
The CR spectrum is discretized using either $N_{\rm c}=5$ or $10$ logarithmically spaced momentum bins between $p_{\rm min}=100\,{\rm MeV}/c$ and $p_{\rm max}=1\,{\rm TeV}/c$.
We adopt a reduced speed of light $\tilde c=6\times10^4\,\rm km\,s^{-1}$, chosen to be large enough that it does not affect the propagation of the highest-momentum CR bins.
The magnetic field is initialized as a tangled field with random orientation, mean strength $B_0$, and coherence length of $10\,\rm pc$, generated following the procedure of \citet{Dubois19} to ensure $\nabla.\vec B=0$.
The magnetic field is sufficiently weak that it does not directly influence the gas dynamics, but it sets the direction of anisotropic CR diffusion.
The mesh is initialized with a uniform grid of $64^3$ cells (corresponding to level $\ell_{\rm min}=6$ and a maximum cell size $\Delta x_{\rm max}\simeq 3.1\,\rm pc$), and is allowed to refine and derefine dynamically up to level $\ell_{\rm max}=8$ ($\Delta x_{\rm min}\simeq 0.8\,\rm pc$).
Refinement is triggered when the relative cell-to-cell variation of gas density or pressure exceeds $10\,\%$, ensuring adequate resolution of the shock, shell, and CR gradients.

The SN explosion is initialized as a spherical region of radius $5\,\rm pc$ with a total energy of $10^{51}\,\rm erg$.
A fraction of $10\,\%$ of this energy is injected as CR energy~\citep{Morlino12,Dermer13,Caprioli14}, following the same initial power-law slope $q_{\rm ini}=4.5$ as the background CRs, while the remaining $90\,\%$ is injected as thermal gas energy.

The gas is allowed to cool radiatively down to $T=T_0$, assuming a metallicity $Z=0.002$ with solar relative elemental abundances.
We use the cooling rates of \citet{Sutherland93} above $10^4\,\rm K$ and of \citet{Dalgarno72} below $10^4\,\rm K$.
CRs lose energy through Coulomb and hadronic interactions.
For simplicity, we neglect ionization losses and assume that the gas is fully ionized at all temperatures\footnote{This assumption slightly underestimates low-energy losses but avoids introducing additional microphysical complexity in this demonstration.}, including at $T=100\,\rm K$.
By default, CRs are allowed to diffuse anisotropically along magnetic field lines, with no explicit perpendicular diffusion. 
In Appendix~\ref{appendix:isotropic_diffusion_SNR}, we also present a simulation with isotropic diffusion to illustrate the flexibility of our solver.
The parallel diffusion coefficient follows a momentum-dependent scaling $\kappa(p) = \kappa_0 (pc/1\rm GeV)^{0.5}$. 
We explore different values of the normalization $\kappa_0$ (see Table~\ref{tab:snr_sim}) in the range $\kappa_0=3\times10^{25}\mbox{--}3\times 10^{26}\,\rm cm^2\,s^{-1}$, as well as a run with virtually no diffusion ($\kappa_0=10^{20}\,\rm cm^2\,s^{-1}$).
These diffusion coefficients are lower than the canonical mean Galactic value~\citep[e.g.][]{Evoli20}, but are motivated by the efficient self-excitation of Alfv\'en waves expected near CR sources~\citep{Nava16,Brahimi20}. 
A more complete model would self-consistently account for the modulation of diffusion by such processes, leading to spatial and temporal variations of $\kappa$, but this is beyond the scope of the present paper and will be addressed in future work.

Table~\ref{tab:snr_sim} summarizes all the SNR simulations analyzed in this section.
Unless stated otherwise, we focus below on the reference run SNRD26.

Figure~\ref{fig:sedD26} shows thin-slice maps of the CR energy distribution at $t=200 \,\rm kyr$ for a random magnetic field with $B_0=0.025\, \rm \mu G$ and $\kappa_0=3\times10^{26} \,\rm cm^2\,s^{-1}$. 
Each panel displays the CR energy density for one of the $N_{\rm c}=5$ momentum bins, with the last panel showing the total CR energy density.
Although CRs diffuse away from the remnant, most of the CR energy remains concentrated near the center, within the bubble and around the shell. 
The energy budget is dominated by CR groups with momenta $1$ and $10\, {\rm GeV}/c$. 
Higher-momentum CRs are progressively less confined, and at TeV energies the distribution becomes nearly uniform across the computational domain.
Outside the SNR, CRs form narrow plumes emerging from the shell edge, tracing magnetic flux tubes and reflecting anisotropic diffusion along field lines~\citep[e.g.][]{Girichidis14}.

\begin{figure}
    \centering
    \includegraphics[width=\columnwidth]{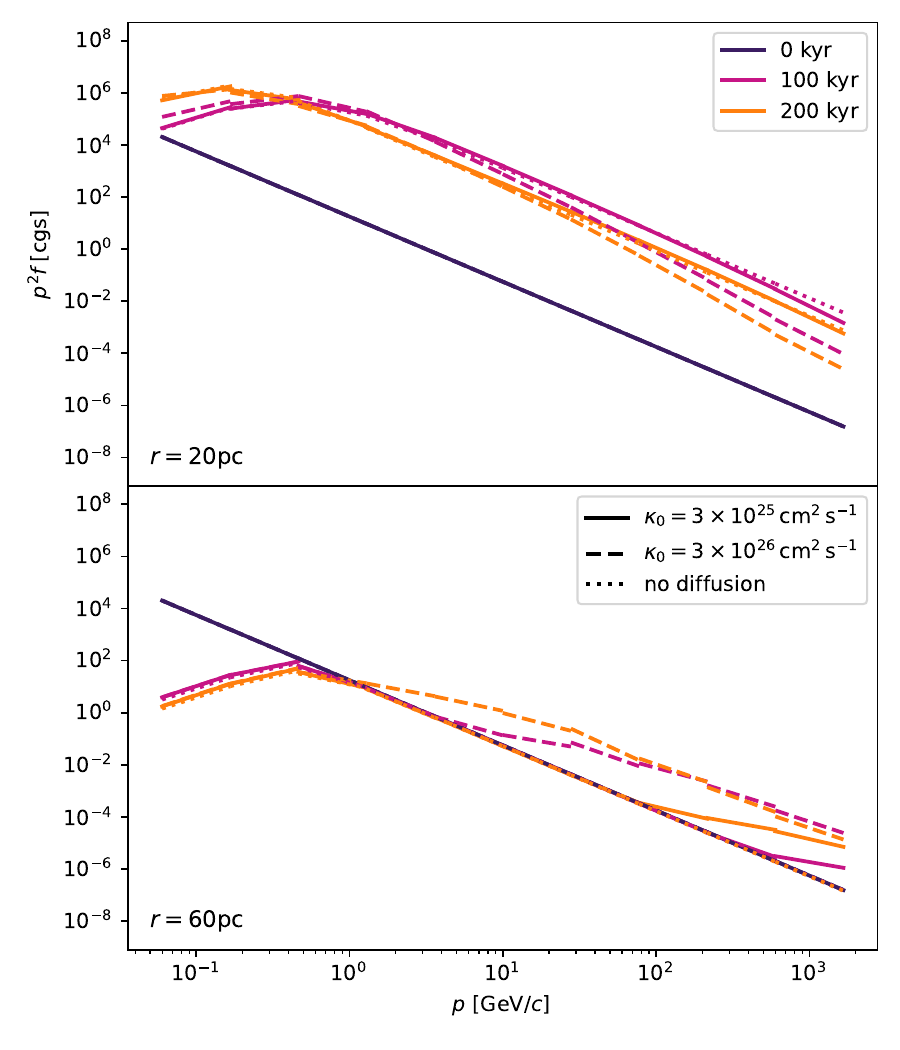}
    \caption{Time evolution of the CR distribution function in SNR simulations. Each panel shows the CR spectrum at a different position. Within a panel, the spectrum is shown at three different times for simulations using different diffusion coefficients (SNRD25N10, SNRD26N10 and SNRnodiff see Table~\ref{tab:snr_sim}). We follow $N_{\rm c}=10$ momentum bins, initialized with $f \propto p^{-4.5}$, $n=10\,\rm cm^{-3}$ and $B_0=0.025 \mu \rm G$. The spectral evolution depends on the measurement location.}
    \label{fig:spectra_snr_allkappa}
\end{figure}

To further characterize CR transport and highlight a key capability of {\sc ramses-mcr}, we examine the CR distribution function at different distances from the explosion.
Figure~\ref{fig:spectra_snr_allkappa} shows the time evolution of the CR spectrum at two radial positions (averaged over spherical shells of width $4\,\rm pc$) for the runs SNRD25N10, SNRD26N10 and SNRnodiff.
We consider here simulations with $N_{\rm c}=10$ momentum bins and verified that reducing the number of bins to $N_{\rm c}=5$ does not change the results.
The top panel corresponds to the shell position ($20\,\rm pc$ at $t=200 \,\rm kyr$, $15\,\rm pc$ at $100\,\rm kyr$), while the bottom panel probes the upstream region affected by diffusive CR leakage.
The spectra reflect the combined effects of radiative cooling, adiabatic compression or expansion, and diffusion. 
All locations exhibit signatures of Coulomb losses at low momenta ($p\lesssim1\,{\rm GeV}/c$), visible as a bending of the distribution function.
At $20\, \rm pc$, the high-momentum component rises first, as higher-energy CRs reach this region earlier owing to their larger diffusion coefficient. 
When the shock arrives, adiabatic compression amplifies the spectrum.
At $60\, \rm pc$, background low-momentum CRs cool and decrease in amplitude, while the high-momentum component grows as progressively more SN-accelerated CRs diffuse outward, again with the highest momenta arriving first.

\begin{figure}
    \centering
    \includegraphics[width=\columnwidth]{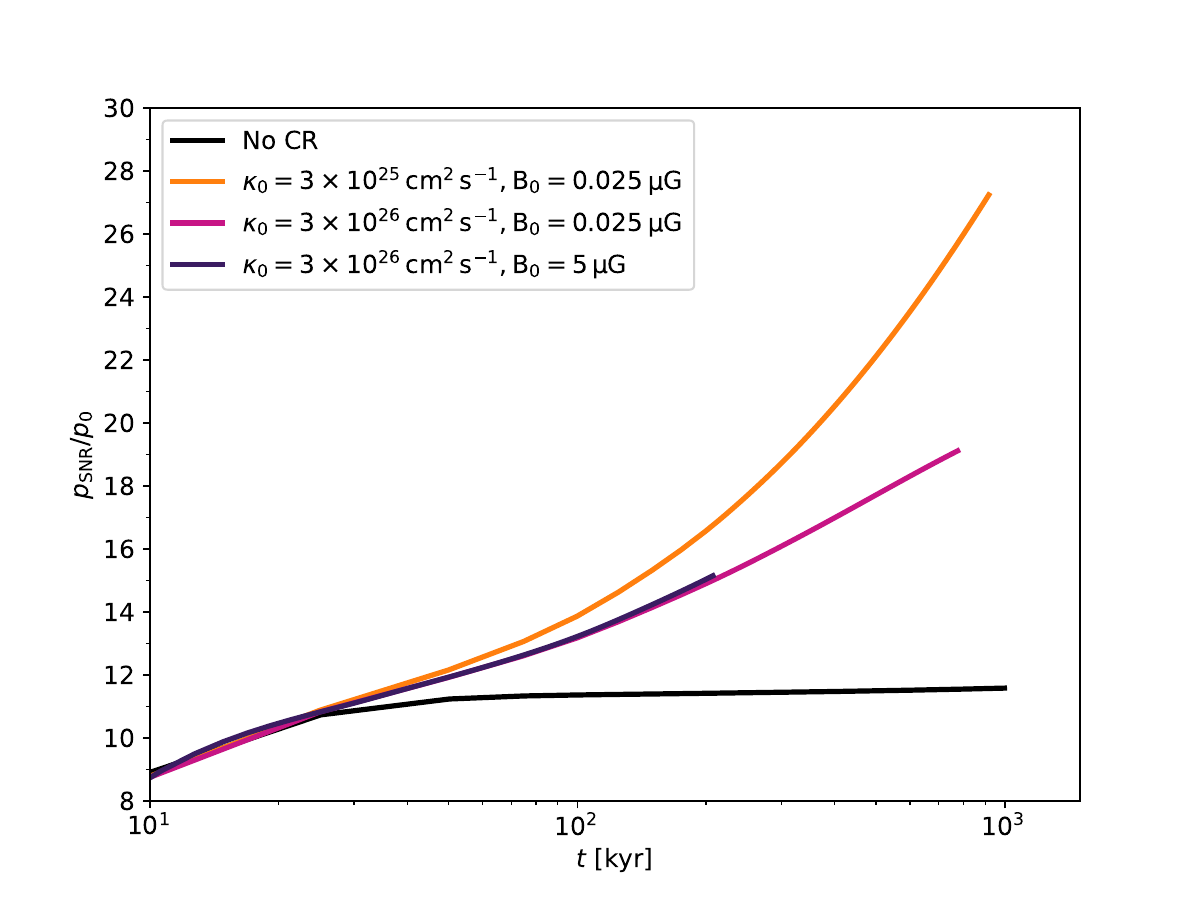}
    \caption{Evolution of the normalized radial gas momentum of the SNR as a function of time for the different simulations. The black line corresponds to the run without CRs, while the three other lines show the results from the CRMHD simulations. The evolution of $p_{\rm SNR}$ is strongly affected by the presence of CRs and their transport properties. The largest enhancement is obtained for the lowest diffusion coefficient, reaching $135\%$ with respect to the no-CR run.}
    \label{fig:mom_snr}
\end{figure}

We next examine the time evolution of the SNR radial momentum for the different simulations. 
Figure~\ref{fig:mom_snr} shows the momentum $p_{\rm SNR}$ measured within a sphere of radius $50 \, \rm pc$, encompassing both the bubble and the shell, as a function of time.
The momentum is normalized to $p_0=\sqrt{2 m_{\rm ej} E_{\rm SNR}}$, adopting $m_{\rm ej}=2 \,\rm M_\odot$ and $E_{\rm SNR}=10^{51}\, \rm erg$ following~\citealp{Rodriguez22}. 
The evolution is shown from $10\, \rm kyr$ (just before the end of the Sedov–Taylor phase, \citealp{Kim15}) to $1\, \rm Myr$.
The black curve corresponds to the run without CRs and serves as a reference. 
In this case, the momentum increases until $\sim 50\,\rm kyr$, after which it reaches a plateau. 
This behavior is expected for SNRs expanding in a uniform medium: once radiative cooling becomes important, the remnant transitions from the pressure-driven snowplough phase to the momentum-conserving phase.

We now turn to the CRMHD simulations. 
In the presence of CRs, the SNR momentum continues to increase beyond the maximum value of the reference run, exceeding it as early as $\sim 10\,\rm kyr$.
The diffusion coefficient strongly impacts the momentum growth. 
The largest enhancement occurs for the lowest diffusion coefficient, reaching an increase of about $135\%$ at $t\approx0.9 \, \rm Myr$ relative to the reference case. 
For higher diffusion, the enhancement is reduced to about $65\%$ at $t\approx0.8 \, \rm Myr$.
This sensitivity arises because diffusion controls the residence time of CRs within the remnant. 
As the SNR expands, the thermal energy density decreases more rapidly than the CR energy density, leading CRs to dominate the internal energy budget at late times. 
This delays the transition to the momentum plateau and enhances the additional momentum growth driven by CR pressure.
Finally, increasing the magnetic field strength (run SNRD26B5) does not significantly modify the momentum evolution.

These results highlight that the dynamical impact of CRs is intrinsically energy-dependent. 
Because the diffusion coefficient increases with particle momentum, high-energy CRs escape rapidly from the SNR, whereas lower-energy CRs remain confined longer and sustain the internal pressure support. 
The resulting momentum boost reflects the time-dependent contribution of spectral components with different transport and loss timescales.

\begin{table}[ht]
\centering

\begin{tabular}{lcccccc}

\hline
Simulation & $B_0\,[\mu\mathrm{G}]$ & $\kappa_0\,[\mathrm{cm}^2\,\mathrm{s}^{-1}]$ 
& $N_{\rm c}$ \\
\hline
NoCR       & $0.025$   & / & /  \\
SNRnodiff   & $0.025$   & $1\times10^{20}$  & 10  \\
SNRiso   & $0.025$   & $1\times10^{26}$  & 5  \\
SNRD25      & $0.025$ & $3\times10^{25}$ & 5  \\
SNRD26      & $0.025$ & $3\times10^{26}$ & 5  \\
SNRD26B5    & $5$     & $3\times10^{26}$ & 5  \\
SNRD25N10   & $0.025$ & $3\times10^{25}$ & 10  \\
SNRD26N10   & $0.025$ & $3\times10^{26}$ & 10  \\
\hline
\end{tabular}
\caption{Summary of all the SNR simulations. For all simulation listed here, the box size is $200\, \rm pc$. They all have a minimum number of cell of $64^3$ and maximum of $256^3$ cells. For each simulation, columns show: initial magnetic field strength $B_0$, diffusion coefficent normalization $\kappa_0$ at $1 \mathrm{GeV}/c$ and the number of CRs bin $N_{\rm c}$. In every case the magnetic field has a random direction with typical B-field coherence length of $10 \rm pc$.}
\label{tab:snr_sim}
\end{table}

\section{Conclusions}
\label{section:conclusion}

We have presented in this paper {\sc ramses-mcr}, the implementation of a  multi-group CR method for the AMR code {\sc ramses}~\citep{Teyssier02}, that is an extension in the spectral domain of two-moment CR method introduced in~\Rosdahl.
The numerical method assumes that the CR distribution function is a power-law in momentum space~\citep{Jones99, Miniati01}, that requires to track both the CR energy and number densities for multiple CR groups.
The method relies on a discretization in momentum space and, hence, enables momentum-dependent anisotropic diffusion, streaming, and radiative losses.

We performed a series of tests for {\sc ramses-mcr} to verify the robustness of the particular numerical implementation of this method.
The distortion of the CR distribution function in the non-relativistic and relativistic due to radiative losses by respectively Coulomb and hadronic losses is correctly captured, with consistent re-channeling of theses losses in the thermal pool (Section~\ref{section:freecooling}).
We tested the injection of fresh CRs and the evolution of the distribution towards steady state with radiative losses counter-balancing the injection, and the analytical expectation is correctly reproduced (Section~\ref{section:steady}).
We tested the impact of the time integration scheme using an adiabatic compression test, which shows improved results for the semi-implicit scheme, with errors of only a few percent (Section~\ref{section:adiabatic}).
We simulated the one-dimensional spatial diffusion of CRs with a momentum-dependent scaling law, and we accurately capture the deformation of the CR distribution function with inner slopes correctly connecting with its global slope (Section~\ref{section:1d_diffusion}), which is extended to verifying the correct behavior of the multi-group anisotropic diffusion within a multi-dimensional setup (Section~\ref{section:2d_diffusion}).
We simulated a shock tube validating the effect of CR streaming with the multi-group approach for spectral CRs (Section\ref{section:shock_tube}).

We have shown that {\sc ramses-mcr} accurately reproduces the theoretical CR steady state spectra under radiative cooling and injection and correctly captures multi-group momentum diffusion. 
The method is numerically stable, as verified by tests of the time integration scheme in the case of adiabatic compression, which show only small errors for the semi-implicit scheme.
For the one-dimensional Sod shock tube test without streaming, our results are consistent with those obtained using the ``grey'' (mono-group)  method of \Rosdahl.

Finally we used {\sc ramses-mcr} to perform a CRMHD simulation of a SNR and we explored different diffusion coefficient normalizations ($3\times10^{25}$ and $3\times10^{26} \,\rm cm^2\, s^{-1} $) and magnetic field strengths ($0.025$ and $5 \,\mu\rm G$).
In agreement with previous studies \citep{Diesing18,Rodriguez22}, we find the emergence of an additional CR pressure–driven snowplough phase, during which the SNR momentum continues to increase.

Overall, {\sc ramses-mcr} provides a robust tool for studying CR feedback using a more accurate treatment of CR physics.
Future applications of this method will include the study of SNRs dynamics with spatially varying CR diffusion coefficients and their associated non-thermal emission, and the impact of CRs on wind formation and regulation of the cold gas in galaxies.

\section*{Acknowledgements}
{We thank Fabio Acero, Ludwig B\"oss, Philippe Girichidis, Isabelle Grenier, Arturo Núñez-Castiñeyra, Francisco Rodríguez Montero, Marco Padovani, and Romain Teyssier for stimulating discussions. This work has made use of the Infinity Cluster hosted by Institut d'Astrophysique de Paris; we thank St\'ephane Rouberol for running smoothly this cluster for us. For the purpose of open access, the author has applied a Creative Commons Attribution (CC BY) licence to any Author Accepted Manuscript version arising from this submission. This work was supported by the``Action Thématiques'', ``Phénomènes Extrêmes et Multi-messagers'' (ATPEM) and ``Physique Chimie du Milieu Interstellaire'' (AT-PCMI) of CNRS/INSU.}


\appendix

\section{The two-moment method limited by the reduced speed of light}
\label{appendix:frame_comp}

\begin{figure*}
    \centering
    \includegraphics[width=0.9\columnwidth]{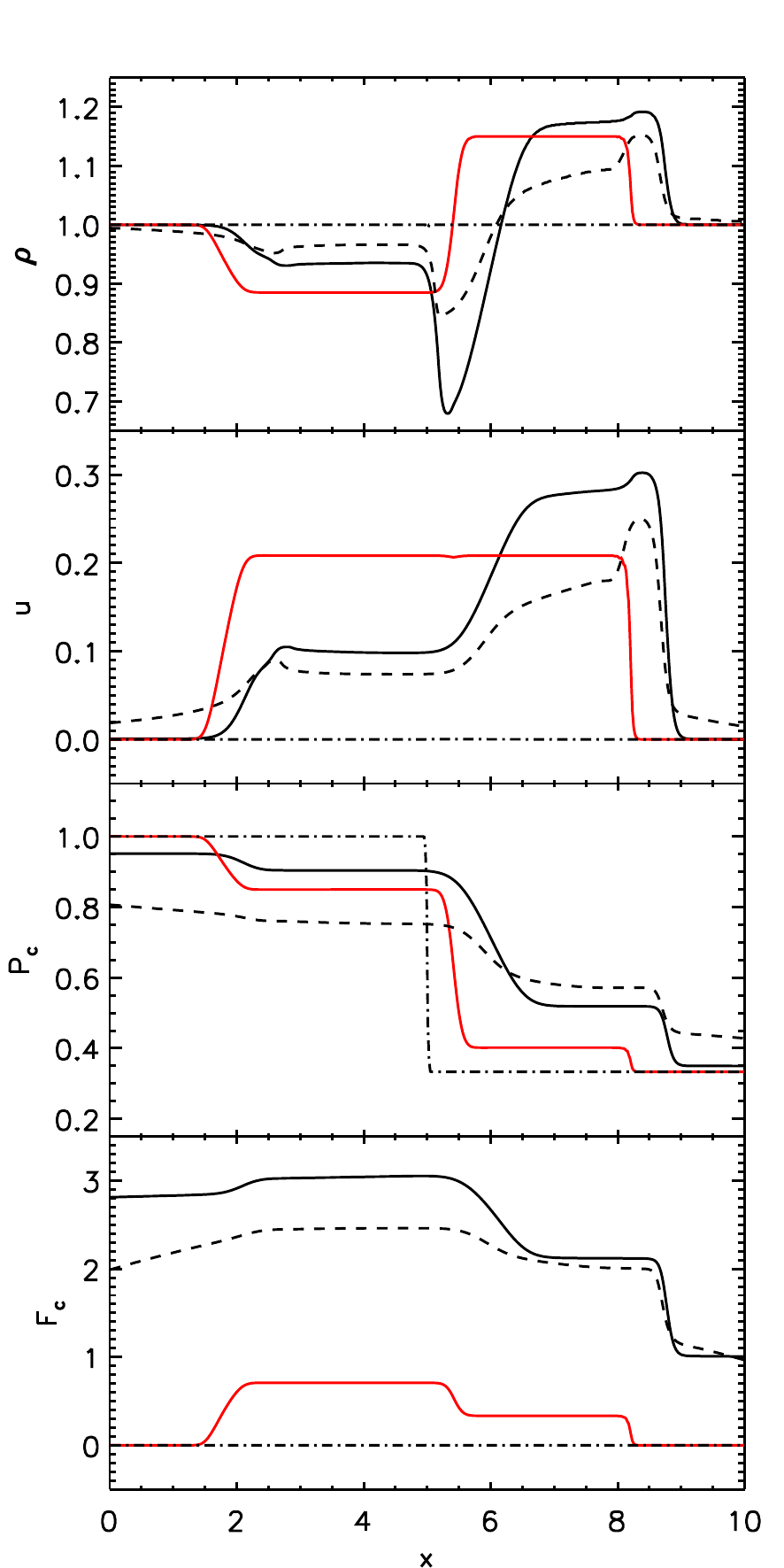}
    \includegraphics[width=0.9\columnwidth]{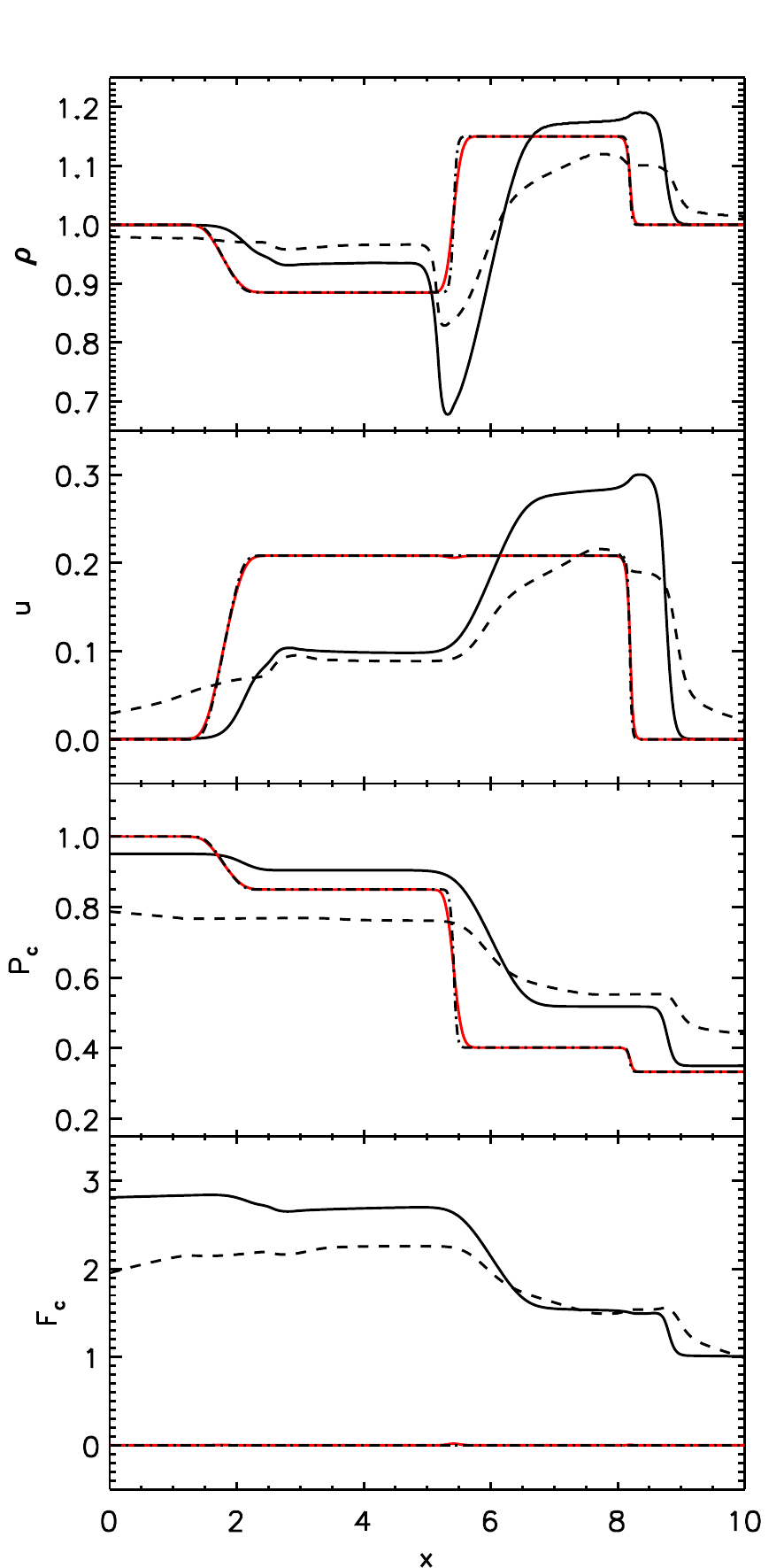}
    \caption{Shock tube test dominated by advection and streaming with imposed constant Alfv\'en velocity of $0.75$ (black lines) showing the result at $t=2$ for the lab-frame formulation (left panels) and the comoving fluid-frame formulation adopted in this work (right panels) using different values of the reduced speed of light $\tilde c=100$ (solid), $\tilde c=1$ (dashed), and $\tilde c=0.01$ (dot-dashed). For comparison, we show the result of the simulation that does not include streaming (red lines).}
    \label{fig:shock_tube_frame}
\end{figure*}

The two-moment formalism limits the propagation of CRs to the speed of light $c$.
In practical applications, a reduced value of the speed of light (denoted $\tilde c$ in the equations above) is adopted. This value must remain sufficiently large compared to any other characteristic wave speed in the system (e.g.~Alfvén velocity, diffusion velocity) so that the limit $\tilde c \rightarrow c$ is well approximated.
However, in complex systems this condition is not always satisfied, for instance due to the evolution of the Alfvén velocity or the use of a variable diffusion coefficient.

The lab-frame and comoving fluid-frame formulations differ in an important way: in the lab-frame, the advection of the CR flux by the gas is also limited by $\tilde c$, whereas this is not the case in the fluid-frame formulation. It is therefore necessary to test the behavior of the two-moment method in regimes where CR transport is effectively limited by $\tilde c$ (i.e.~far from flux steady state).
In addition, the spectral method requires isolating the contribution of the $-P_{\rm c}\vec\nabla.\vec u$ term at each spectral step to ensure conservation of the slope of the distribution function during adiabatic transformations. This is only possible within the comoving fluid-frame formulation of the two-moment equations.

For clarity, the fluid-frame formulation tested here for a grey spectrum (one bin) is:
\begin{align}
\label{eq:ecr_fluid} 
&    \frac{\partial e_{{\rm c}}}{\partial t} + \vec \nabla . (\vec u e_{{\rm c}}) + \vec \nabla.\vec F^e_{{\rm c}} = - P_{{\rm c}}\vec \nabla . \vec u -\vec u_{{\rm s}}.\sigma^e\vec F^e_{{\rm c}} \, ,\\
&    \label{eq:fecr_fluid} \frac{1}{\tilde c^2}\frac{\partial \vec F^e_{{\rm c}}}{\partial t} + \vec \nabla \left(\frac{e_{{\rm c}}}{3}\right) = -\sigma^e\vec F^e_{{\rm c}}\, ,
\end{align}
while the lab-frame formulation (as in~\Rosdahl) is:
\begin{align}
\label{eq:ecr_fluid} 
&    \frac{\partial e_{{\rm c}}}{\partial t} + \vec \nabla.\vec F^e_{{\rm c,lab}} = \vec u.\vec \nabla P_{{\rm c}} -\vec u_{{\rm s}}.\sigma^e(\vec F^e_{{\rm c,lab}}-\vec u(e_{\rm c}+P_{\rm c})) \, , \\
&    \label{eq:fecr_fluid} \frac{1}{\tilde c^2}\frac{\partial \vec F^e_{{\rm c,lab}}}{\partial t} + \vec \nabla \left(\frac{e_{{\rm c}}}{3}\right) = -\sigma^e(\vec F^e_{{\rm c,lab}}-\vec u(e_{\rm c}+P_{\rm c}))\, .
\end{align}

We set up a shock-tube test similar to the test in~\cite{Thomas21}.
All hydrodynamical quantities in a box of size 10 are initialized to uniform values $\rho=1$, $u=0$, $P=1$, $B=10^{-5}$, and $F^e_{\rm c,lab}=F^e_{\rm c}=0$, except for the CR pressure, which is set to $P_{\rm c}=1$ on the left-hand side and $P_{\rm c}=1/3$ on the right-hand side.
The adiabatic index of the gas is $\gamma=5/3$, and that of the CRs is $\gamma_{\rm c}=4/3$.
We adopt a constant Alfvén velocity of $u_{\rm A}=0.75$ for streaming, and a small diffusion coefficient $\kappa=1/300$.
The simulation is run with a constant resolution of 512 cells.

In Fig.~\ref{fig:shock_tube_frame} we show the simulation results for the gas density, velocity, CR pressure, and CR flux for different reduced speeds of light, $\tilde c=0.01$, $1$, and $100$, in both the lab-frame and comoving fluid-frame formulations.
For comparison, we also show a simulation without streaming at $\tilde c=100$, whose solution at $t=2$ corresponds to a shock wave propagating to the right, a rarefaction wave to the left, and a central contact discontinuity.

With streaming and $\tilde c=100$ (our reference solution), the flux reaches steady state and both the lab-frame and fluid-frame formulations give the same result. The solution qualitatively resembles that without streaming: a left/right-most rarefaction/shock wave propagating at a speed reduced/increased by the streaming velocity. Around the central contact-like discontinuity, streaming transfers CR energy efficiently into the thermal reservoir due to the sharp CR pressure gradient. Because of the different adiabatic indices of the CR and thermal components, the gas develops a stronger pressure gradient from streaming losses, leading to the central depletion.

For an intermediate value $\tilde c=1$, the solution is strongly modified since $\tilde c$ is comparable to the Alfvén velocity ($u_{\rm A}=0.75$) and to the shock velocity ($\simeq 1.5$). The lab- and fluid-frame solutions at this $\tilde c$ still show qualitatively similar structures, but with noticeable differences in shock positions and discontinuities.

For the extremely small value $\tilde c=0.01$, much lower than all characteristic wave speeds, the two formulations diverge completely despite having identical non-evolving fluxes. In the lab-frame formulation, the solution at $t=2$ remains essentially unchanged from the initial conditions, since the low $\tilde c$ suppresses both streaming and advection of CRs by the gas, and thus suppresses the gas motion itself (which would otherwise be driven by the initial CR pressure jump). In contrast, the fluid-frame formulation still captures the gas dynamics, since the convective terms are absent from the flux by construction. The resulting solution at $t=2$ is therefore close to that without streaming (and $\tilde c=100$, in red), apart from missing a slight smearing of the central contact discontinuity at $x=5$ caused by diffusion. This smearing disappears in the $\tilde c=0.01$ case because the effective diffusion velocity at the grid scale is $u_{\rm D}=\kappa/\Delta x\simeq 0.17>\tilde c$.

\section{Cosmic ray equations with appropriate limits}
\label{appendix:extra_equations}

The equations for CR transport are rewritten here in full:
\begin{align}
&    \frac{\partial \nci}{\partial t} + \vec \nabla . (\vec u \nci) +\vec \nabla . (F^\nci\vec b)=\left[4\pi p^2 L(p) f_0\right]_{\pim}^{\pip} + j^n_{0,i}\, ,\nonumber\\
&    \frac{1}{\tilde c^2}\frac{\partial F^\nci}{\partial t} +\vec b.\vec \nabla \left(\frac{\nci}{3}\right) = -\frac{1}{3\kappa^\nci} \left[F^\nci-\frac{q_i}{3} \bar u_{\rm A} \nci \right]\, ,\nonumber\\
&    \frac{\partial \eci}{\partial t} + \vec \nabla . (\vec u \eci) + \vec \nabla.(F^\eci \vec b) =  \left[4\pi p^2 L(p) T(p) f_0(p)\right]_{\pim}^{\pip}\nonumber\\
&+\mathcal{L}_{\rm r}-P_{{\rm c},i}\vec \nabla.\vec u \nonumber
-\frac{(\gci-1)\bar u_{\rm A}}{3\kappa^\eci}\left[F^\eci-\frac{q_i}{3}\bar u_{\rm A}\eci\right] + j^e_{0,i} \, ,\nonumber\\
&    \frac{1}{v^2}\frac{\partial F^\eci}{\partial t} +\vec b .\vec \nabla \left (\frac{\eci}{3} \right)= -\frac{1}{3\kappa^\eci} \left[F^\eci-\frac{q_i}{3}\bar u_{\rm A} \eci\right]\, ,\nonumber
\end{align}
where $\mathcal{L}_{\rm r}$ are the spectrally integrated radiative losses. 
An important limit is when the flux is in local steady state, leading to the  following form of the equations on $\nci$ and $\eci$:
\begin{align}
&\frac{\partial \nci}{\partial t} + \vec \nabla.\left(\left(\vec u+\frac{q_i}{3}\bar u_{\rm A}\right) \nci\right) = - \vec \nabla.(-\kappa^\nci \vec b\vec b.\vec \nabla \nci) + j^n_{0,i}\nonumber\\
&+ \left[4\pi p^2 L(p) f_0(p)\right]_{\pim}^{\pip} \, ,\nonumber\\
&\frac{\partial \eci}{\partial t}+\vec \nabla.\left(\left(\vec u+\frac{q_i}{3}\bar u_{\rm A}\right)\eci\right) = - \vec \nabla.(-\kappa^\eci \vec b \vec b. \vec \nabla \eci) + j^e_{0,i}\nonumber\\
&+\mathcal{L}_{\rm r}-P_{{\rm c},i}\vec \nabla.\vec u + \bar u_{\rm A}.\vec\nabla \pci +\left[4\pi p^2 L(p) T(p) f_0(p)\right]_{\pim}^{\pip}\, .\nonumber
\end{align}
The grey 1-moment limit is obtained by closing the $p$-bin boundaries with $\pim\rightarrow 0$ and $\pip\rightarrow\infty$ (i.e.~$[...]_{\pim}^{\pip}=0$). 
In addition, assuming $q_i=4$ (which leads to dropping the equation on $\nc$), this reduces to a single equation on energy:
\begin{align}
&\frac{\partial \ec}{\partial t}+\vec \nabla.\left(\left(\vec u+\frac{4}{3}\bar u_{\rm A}\right)\ec\right) = - \vec \nabla.(-\kappa^\ec \vec b \vec b. \vec \nabla \ec) + j^e_{0}\nonumber\\
&+\mathcal{L}_{\rm r}-P_{\rm c}\vec \nabla.\vec u + \bar u_{\rm A}.\vec\nabla P_{\rm c}\, .\nonumber
\end{align}

\section{Synchrotron losses for cosmic ray electrons}
\label{appendix:cr_electrons}

We present here the result of {\sc ramses-mcr} applied to CR electrons.
When we consider CR electrons, the radiative loss term contains synchrotron losses and inverse-Compton scattering (IC) $L_{\rm r}=L_{\rm r,s} + L_{\rm r,IC}$.
These two processes scale with $p^2$, therefore because the scaling is the same, we neglect the IC scattering for this test ($L_{\rm r,IC}=0$), and only use  synchrotron losses, which loss rate is: 
\begin{equation}
\label{eq:synchr_lossrate}
    L_{\rm r,s}= \frac{4}{3} \frac{\sigma_{\rm T}}{m_{\rm e}^2c^2}\frac{B^2}{8\pi}p^2=\beta p^2\, ,
\end{equation}
where $\sigma_{\rm T}$ is the Thomson cross-section and $m_{\rm e}$ is the mass of the electron.
We also change the rest-mass of the CR particle to that of electrons, which for this particular range of energies ($10\,\rm GeV/c$ to $100\,\rm TeV/c$) capture electrons in their ultra-relativistic regime.

\begin{figure}
    \centering
    \includegraphics[width=\columnwidth]{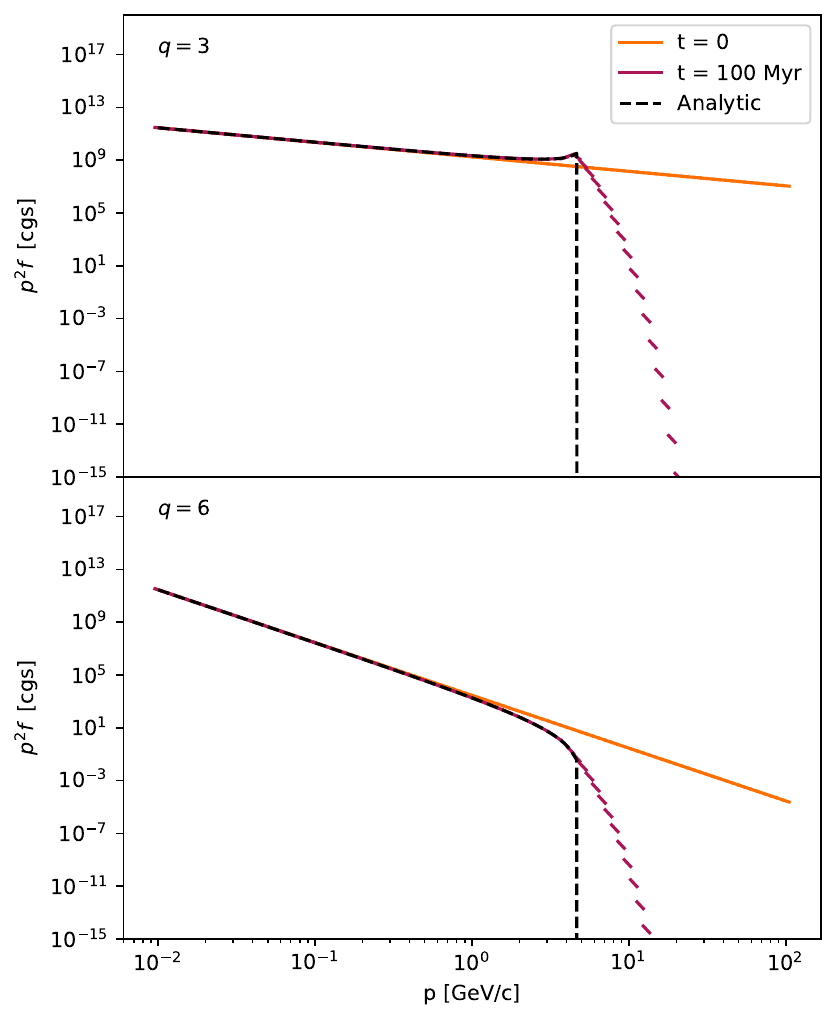}
    \caption{CR electron spectrum evolution cooled by synchrotron losses, with $N_{\rm c}=100$ and a magnetic field strength $B=5\,\mu \rm G$. The first panel uses an initial function distribution $f \propto p^{-3}$ and the second panel $f \propto p^{-6}$. In both panels, the black dashed line indicates the analytic solution at the final time. We obtain a good agreement with the analytic solution for the two spectral shape.}
    \label{fig:synchrotron}
\end{figure}

The analytical solution is that of Appendix~\ref{appendix:solution_synchrotron}.
This solution indicates a difference in spectral shape for spectra with an initial slope of $q<4$ and $q>4$ and the existence of a cutoff at $p_{\rm th}=1/(\beta t)$.
Figure~\ref{fig:synchrotron} shows the result of this test of cooling for CR electrons for two different initial slopes, both assuming a magnetic field strength of $B=5\,\rm \mu G$. 
We find a cooled spectrum in good agreement with the analytic prediction. 
The difference appearing after the cutoff is due to the interpolation of the slope in our solver reaching the maximum of the tabulated values of $q_{\rm max}=10$: using larger values of $q_{\rm max}$ would lead to a sharper drop of the distribution function after $p_{\rm th}$.

\section{Analytical solutions for the distribution function}
\label{appendix:analytical_solution}

We derive here analytical solutions for the distribution function as a function of momentum $p$ at time $t$, i.e.~$f(p,t)$, considering Coulomb and (also ionization) losses, adiabatic changes (also hadronic or streaming), and synchrotron (also inverse Compton) radiation.
We assume an initial power-law distribution of the form
\begin{equation}
f(p,t=0)=f_{\rm ini}(p)=f_0\left(\frac{p}{p_0}\right)^{-q}\,.
\end{equation}

The momentum loss term can be written as:
\begin{align}
\frac{\partial f}{\partial t}
= - p^{-2} \frac{\partial}{\partial p}\left(p^2\frac{{\rm d}p}{{\rm d}t} \right)f
-\frac{{\rm d} p}{{\rm d} t}\frac{\partial f}{\partial p}\,.
\end{align}
This shows that the evolution of $f$ along the characteristic curves $p(t)$ is governed by:
\begin{equation}
\frac{{\rm d} f}{{\rm d} t}=- p^{-2} \frac{\partial}{\partial p}\left(p^2\frac{{\rm d}p}{{\rm d}t} \right)f\,.
\end{equation}
For a momentum loss rate of the generic power-law form ${\rm d}p/{\rm d}t=-\beta p^{\alpha}$ (with $\beta>0$ corresponding to losses), this reduces to
\begin{equation}
\frac{{\rm d} f}{{\rm d} t}=\beta (2+\alpha)p^{\alpha-1}f.
\end{equation}

\subsection{Coulomb losses}
\label{appendix:coulomb_prediction}
For Coulomb or ionization losses, $\alpha=-2$, and therefore ${\rm d}f/{\rm d}t=0$. The distribution function is thus conserved along the characteristic curves $p(t)$. The solution for the momentum evolution is
\begin{equation}
p^3=p_{\rm ini}^3-3\beta t\,.
\end{equation}
Since $f(p,t)=f_{\rm ini}(p_{\rm ini})$, we obtain
\begin{equation}
f(p,t)=f_0\left(\frac{p}{p_0}\right)^{-q}\left(1+\frac{p_{\rm cut}^3}{p^3}\right)^{-q/3}\,,
\end{equation}
where $p_{\rm cut}=(3\beta t)^{1/3}$. For $p\gg p_{\rm cut}$, the solution is identical to the initial distribution. In the opposite limit $p\ll p_{\rm cut}$, the distribution tends to $f_0(p_{\rm cut}/p_0)^{-q}$ and thus becomes independent of $p$.

\subsection{Adiabatic changes, and hadronic and streaming losses}
\label{appendix:hadronic_prediction}

For adiabatic changes (the following also applies to hadronic and streaming losses), the momentum loss rate is linear in $p$, corresponding to $\alpha=1$.
The momentum evolution is then given by:
\begin{equation}
p=p_{\rm ini}\exp(-\beta t)\,.
\end{equation}
The corresponding evolution of the distribution function is $f(p,t)=f_{\rm ini}(p_{\rm ini})\exp(3\beta t)$, which yields to:
\begin{equation}
f(p,t)=f_0\left(\frac{p}{p_0}\right)^{-q}\exp\bigl((3-q)\beta t\bigr)\,.
\end{equation}
Therefore, under a linear momentum loss (or gain) rate, the shape of the distribution function remains identical to the initial power law, with a simple exponential rescaling of its normalization. For a given value of $p$, the distribution function decreases for losses ($\beta >0$) if $q>3$ and increases if $q<3$, and vice versa for gains ($\beta <0$).

\subsection{Synchrotron losses}
\label{appendix:solution_synchrotron}

For synchrotron and inverse Compton losses of CR electrons, the momentum loss rate is quadratic in $p$, corresponding to $\alpha=2$. The momentum evolution is then:
\begin{equation}
p=\frac{p_{\rm ini}}{1+\beta p_{\rm ini}t}\,.
\end{equation}
The corresponding evolution of the distribution function is $f(p,t)=f_{\rm ini}(p_{\rm ini})(1+\beta p_{\rm ini}t)^4$, which leads to:
\begin{equation}
f(p,t)=f_0\left(\frac{p}{p_0}\right)^{-q}\left(1-\frac{p}{p_{\rm th}}\right)^{q-4},
\end{equation}
valid for $p<p_{\rm th}$, with $p_{\rm th}=1/(\beta t)$, and $f(p,t)=0$ otherwise. For $q>4$, the distribution function decreases for $p<p_{\rm th}$, while for $q<4$ it increases, leading to a sharp accumulation of CRs just below $p_{\rm th}$.

\begin{figure}
    \centering
    \includegraphics[width=\columnwidth]{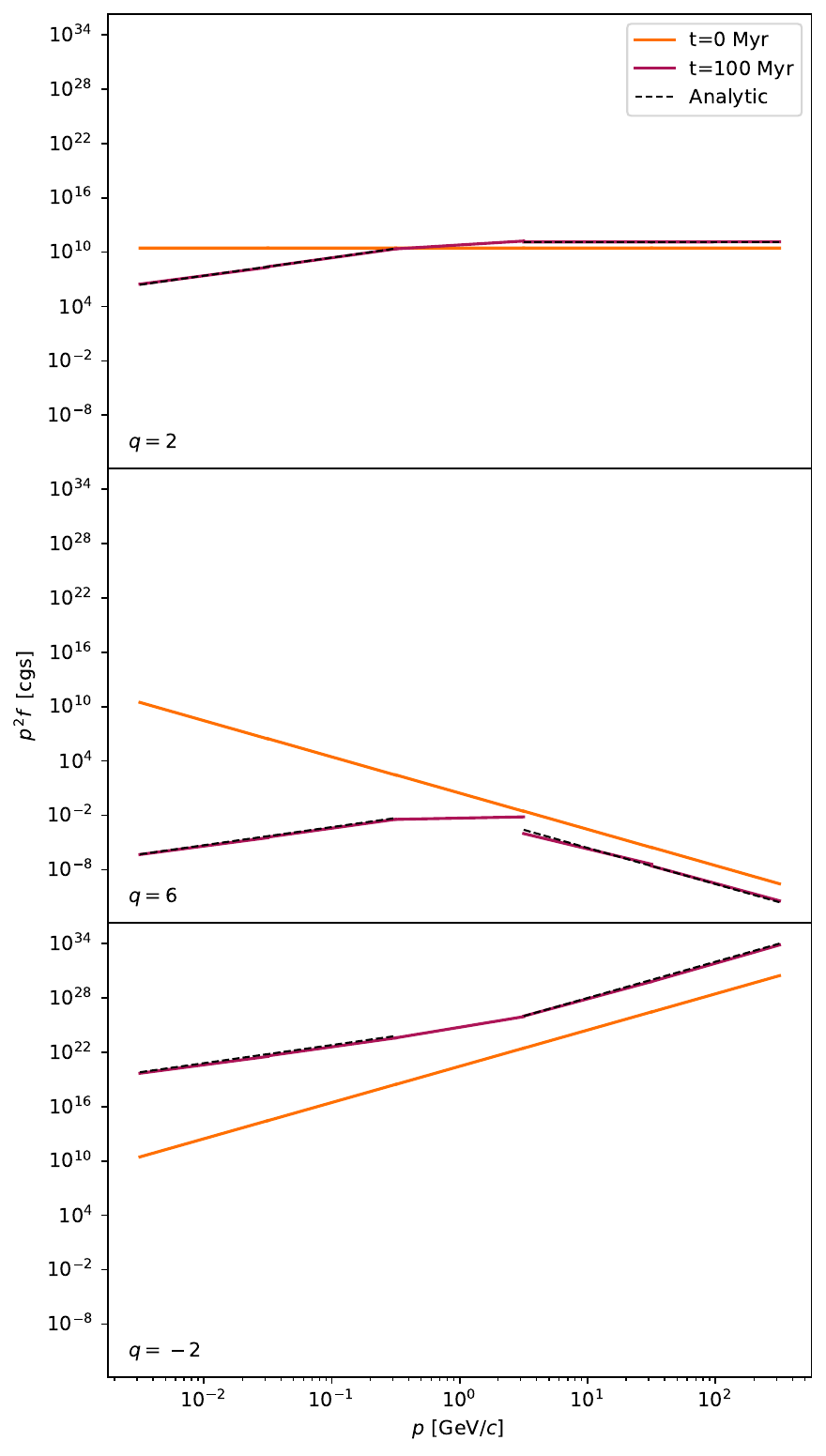}
    \caption{Radiative cooling test, with $N_{\rm c}=5$ number of bins and $n_{\rm e}=1\,\rm cm^{-3}$. The difference between the panels is the initial slope of the distribution function.
    }
    \label{fig:freecool_q2q6}
\end{figure}

\section{Free-cooling as a function of the initial slope}
\label{appendix:freecooling_slope}

Figure~\ref{fig:freecool_q2q6} illustrates that the choice of the initial slope has an effect on the evolution of the distribution function.
In this figure, we do the same free cooling test as in section~\ref{section:freecooling} but for different slopes of the initial distribution function $q_{\rm ini}=2$ (top panel) and $q_{\rm ini}=6$ (middle panel).
As for $q_{\rm ini}=4$, the final slopes of the distribution function are $q=0$ and $q=q_{\rm ini}$ for the sub-relativistic and relativistic regimes respectively as expected for Coulomb losses (see Appendix~\ref{appendix:coulomb_prediction}) and hadronic losses (see Appendix~\ref{appendix:hadronic_prediction}).
However, for $q_{\rm ini}=2$, the distribution function in the relativistic regime has increased instead of decreasing with respect to its initial value as expected for any $q<3$.
This behaviour does not correspond to a physical gain or loss of particles, but rather to a redistribution in momentum space induced by continuous energy losses.
For hadronic losses ($\alpha=1$), the distribution function decreases at fixed momentum for $q>3$, and increases for $q<3$, while conserving the particle number along characteristics in momentum space.
Finally the bottom panel of Fig.~\ref{fig:freecool_q2q6} shows the evolution for $q_{\rm ini}=-2$, where the final solution increases everywhere, i.e.~including the sub-relativistic regime, as expected for Coulomb losses when $q<0$.

\section{Cosmic ray shock tube with various distribution functions}
\label{appendix:shock_streaming_variants}

In this appendix, we present additional shock-tube tests with CR streaming that explore the sensitivity of the solution to the slope of the CR distribution function and the extent of the momentum domain.
These tests are intended to assess the robustness of the spectral formulation and to clarify differences with the grey CR approach, rather than to provide direct one-to-one comparisons.

\begin{figure}
    \centering
    \includegraphics[width=0.49\columnwidth]{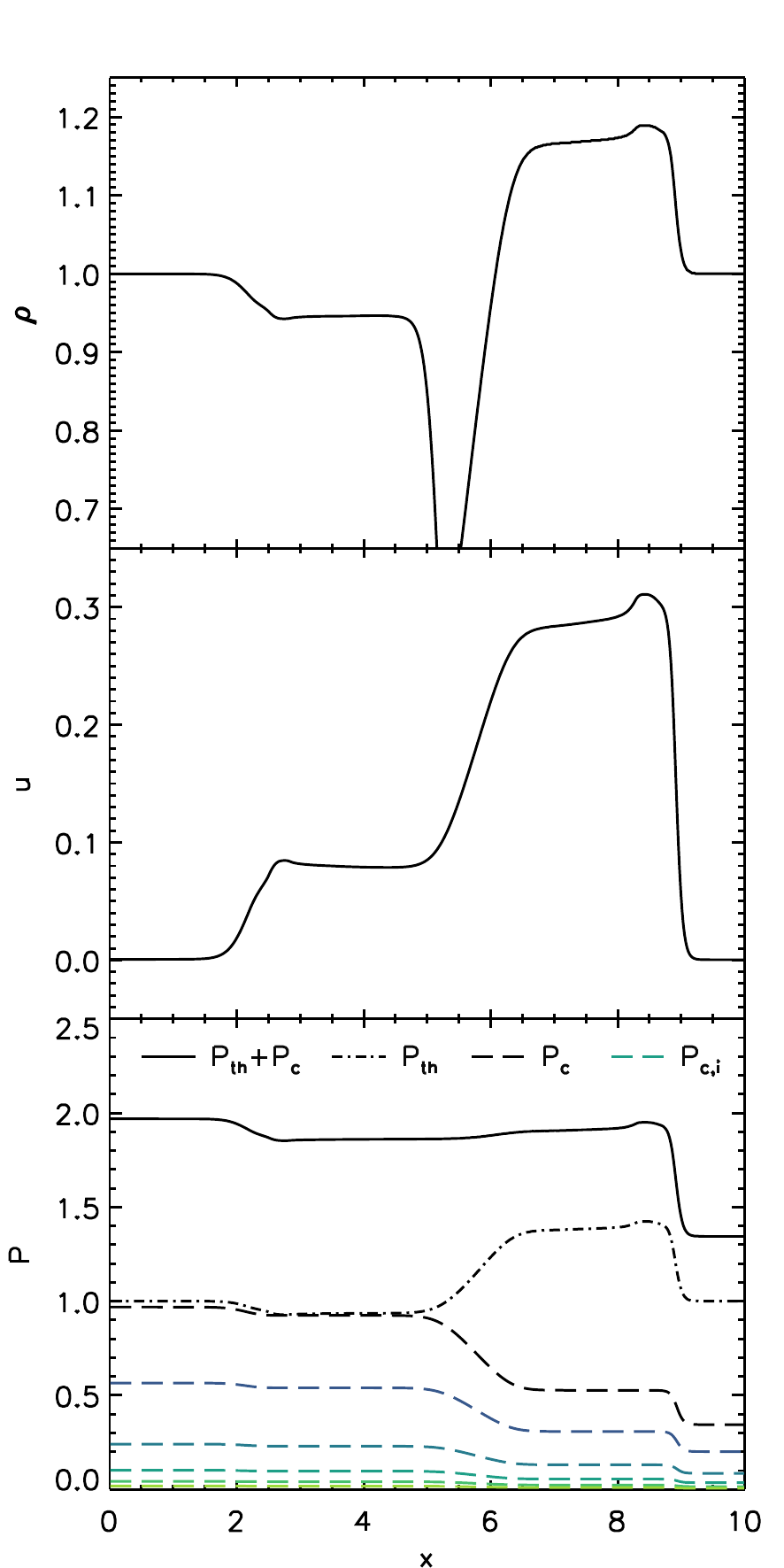}
    \includegraphics[width=0.49\columnwidth]{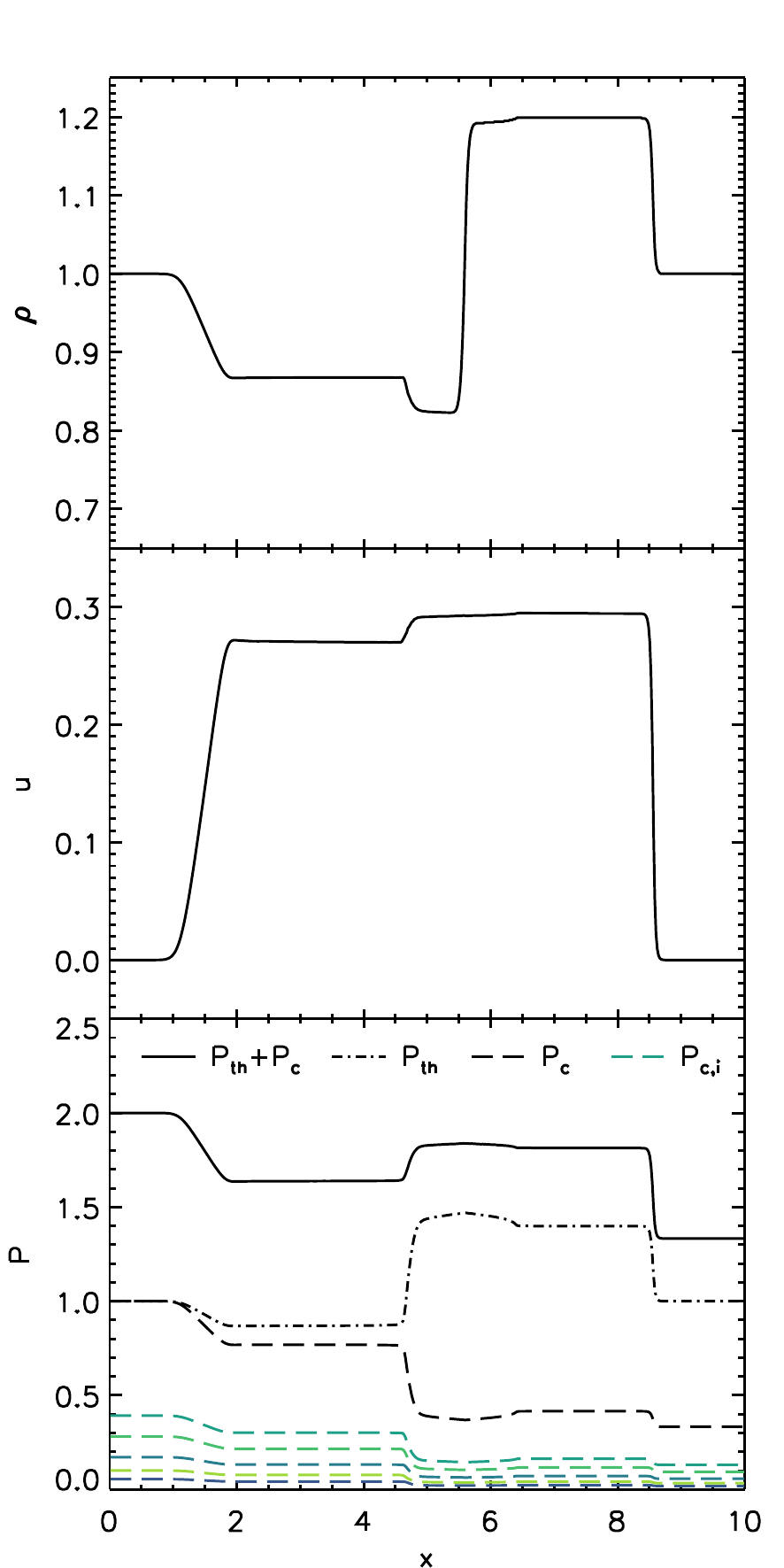}
    \caption{Shock tube test as in the right panel of Fig.~\ref{fig:shock_tube} (i.e.~that includes streaming) except that we have modified the initial distribution function. The left panel shows the solution with $q_{\rm ini}=4.5$ (instead of $q_{\rm ini}=4$) and the right panel shows the solution with $q_{\rm ini}=4.5$ and with the $p$ bins covering $0.01\,{\rm GeV}/c$-$100\,{\rm GeV}/c$ (instead of $100\,{\rm GeV}/c$-$100\,{\rm TeV}/c$).}
    \label{fig:shock_tube_different_fini}
\end{figure}

We performed a test with a different initial slope $q_{\rm ini}=4.5$ instead of the default value of $4$ in Sec.~\ref{section:shock_tube}, which modifies the effective streaming speed.
In this case, shown in left panels of Fig.~\ref{fig:shock_tube_different_fini}, the CR pressure becomes dominated by the lowest momentum bins.
The solution with $q=4.5$ but still with CRs covering the same ultra-relativistic regime show a little bit more depletion in gas density due to larger heating of the gas by streaming.
Even though the losses do not depend directly on $q$ (they are proportional to $\bar u_{\rm A}.\vec \nabla P_{\rm c}$ in flux steady-state), the structure of the shock is modified as CRs move slower or faster by the modified drift speed of CRs, which is directly proportional to $q$.

In a second test (right panels in Fig.~\ref{fig:shock_tube_different_fini}), we shift the sampled CR momentum range to $0.01,{\rm GeV}/c$–$100,{\rm GeV}/c$ in order to include both the sub- and ultra-relativistic regimes.
In this setup, the CR adiabatic index is no longer uniform across momentum and varies between $\gamma_{\rm c}=5/3$ in the sub-relativistic limit and $\gamma_{\rm c}=4/3$ in the ultra-relativistic limit.
Because a significant fraction of the CR energy density now resides in the sub-relativistic regime, where the adiabatic index is larger than $\gamma_{\rm c}>4/3$, the dynamical impact of energy transfer from CRs to the gas through streaming losses is reduced. 
Indeed, the enhanced gas response observed in the ultra-relativistic case relies on the contrast between the CR and gas adiabatic indices ($\gamma_{\rm c}\neq\gamma$).
When $\gamma_{\rm c}$ approaches $\gamma$, the differential pressure response weakens, leading to a smaller effective pressure enhancement and therefore a much less pronounced density depletion at the center.
As a result, the shock structure is qualitatively closer to that obtained in the simulation without CR losses. Only a modest gas density deficit remains at the center, with a $\sim10\%$ variation in velocity across this region (instead of a factor of three difference in the purely ultra-relativistic case).
We also observe that the density and velocity jumps are larger than in the purely ultra-relativistic regime (with or without streaming). This behaviour is consistent with the larger effective equation of state of the CR population and follows from the Rankine–Hugoniot jump conditions.

These additional tests confirm the robustness of {\sc ramses-mcr} and highlight intrinsic differences that cannot be captured by a grey approximation.

\section{Supernova remnant with isotropic diffusion}
\label{appendix:isotropic_diffusion_SNR}

\begin{figure*}
    \centering
    \includegraphics[width=2\columnwidth]{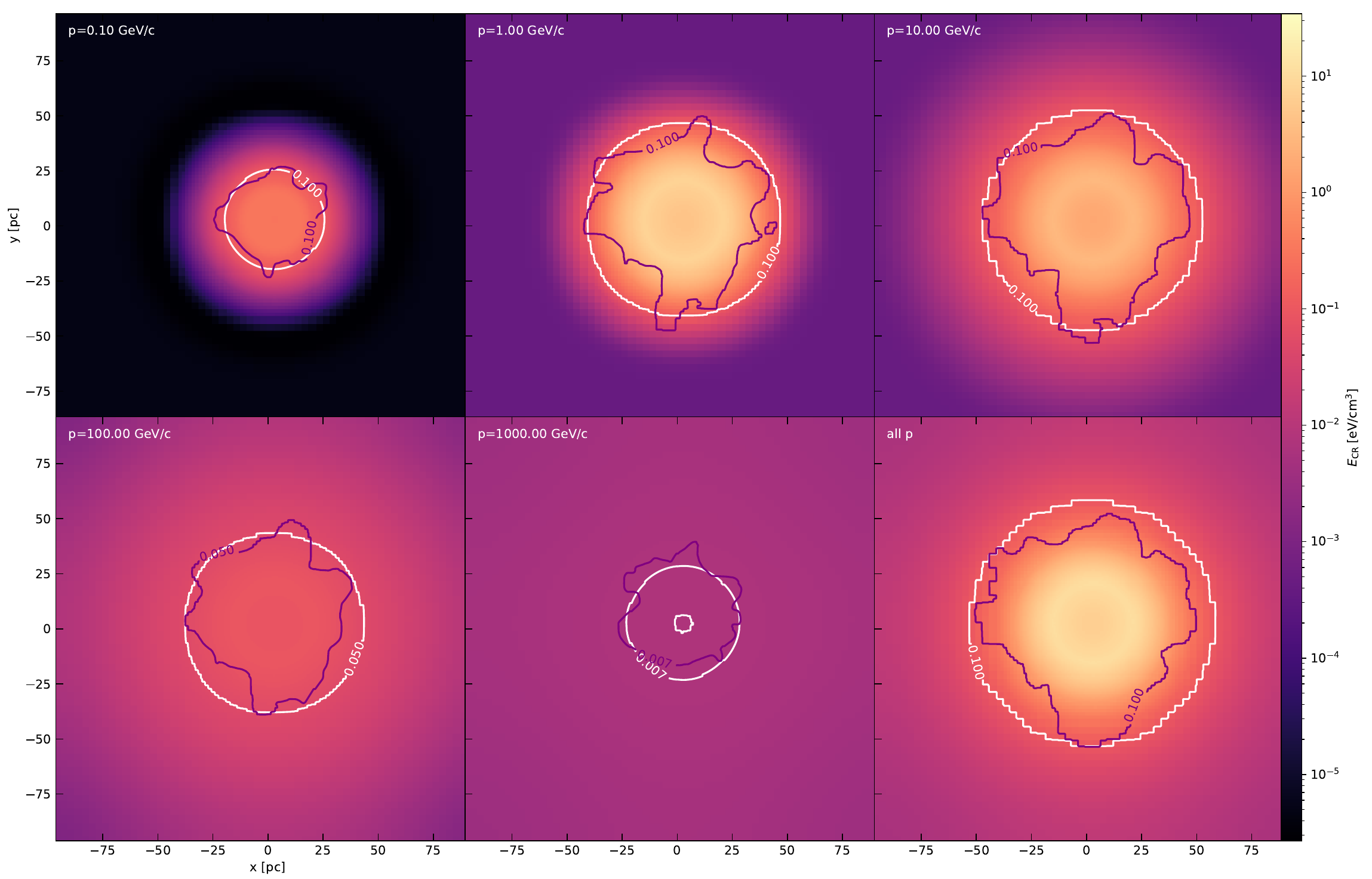}
    \caption{Three-dimensional SN explosion similar to Fig.~\ref{fig:sedD26}, but in the case of isotropic diffusion (SNRiso; see Table~\ref{tab:snr_sim}).
    For comparison, iso-energy contours from the isotropic run are shown in white, while those from the anisotropic case are shown in purple.}
    \label{fig:sediso}
\end{figure*}

To illustrate the versatility of {\sc ramses-mcr}, we performed a SNR simulation with isotropic diffusion (SNRiso), using the same parameters as SNRD26 but assuming $\kappa_0 = 10^{26}\,\rm cm^2\,s^{-1}$.
Figure~\ref{fig:sediso} presents maps of the CR energy distribution from this simulation at $t=200\, \rm kyr$.
The different panels display the CR energy density for each of the $N_{\rm c}=5$ momentum bins, with the last panel showing the total CR energy density.
Iso-energy contours highlighting the difference between isotropic and anisotropic diffusion are overplotted in the figure.
As in the anisotropic diffusion case, most of the CR energy density is concentrated within the bubble and near the shell, and is primarily due to the CRs with momenta $1$ and $10\,\rm GeV/c$.
Compared to Fig.~\ref{fig:sedD26}, CRs are not shaped by the magnetic field but instead  diffuse spherically from the center, with higher-momentum CRs diffusing more rapidly, eventually blending into the background.

\bibliography{references}

\end{document}